\documentclass[aps,prl,twocolumn,superscriptaddress]{revtex4-1}

\usepackage{graphicx}
\usepackage{amssymb}
\usepackage{amsmath}
\usepackage{color}
\begin{document}

\title{Two-gap superconductivity with line nodes in CsCa$_2$Fe$_4$As$_4$F$_2$}

\author{Franziska~K.~K.~Kirschner}
\email{franziska.kirschner@physics.ox.ac.uk}
\affiliation{Department of Physics, University of Oxford, Clarendon Laboratory, Parks Road, Oxford, OX1 3PU, United Kingdom}
\author{Devashibhai~T.~Adroja}
\affiliation{ISIS Facility, Rutherford Appleton Laboratory, Chilton, Didcot Oxon, OX11 0QX, United Kingdom}
\affiliation{Highly Correlated Matter Research Group, Physics Department, University of Johannesburg, PO Box 524, Auckland Park 2006, South Africa}
\author{Zhi-Cheng~Wang}
\affiliation{Department of Physics, Zhejiang University, Hangzhou 310058, China}
\author{Franz~Lang}
\affiliation{Department of Physics, University of Oxford, Clarendon Laboratory, Parks Road, Oxford, OX1 3PU, United Kingdom}
 \author{Michael~Smidman}
\affiliation{Center for Correlated Matter and Department of Physics, Zhejiang University, Hangzhou 310058, China}
\author{Peter~J.~Baker}
\affiliation{ISIS Facility, Rutherford Appleton Laboratory, Chilton, Didcot Oxon, OX11 0QX, United Kingdom}
\author{Guang-Han~Cao}
  \affiliation{Department of Physics, Zhejiang University, Hangzhou 310058, China}
\author{Stephen J.~Blundell}
\email{stephen.blundell@physics.ox.ac.uk}
\affiliation{Department of Physics, University of Oxford, Clarendon Laboratory, Parks Road, Oxford, OX1 3PU, United Kingdom}
\date{\today}

\begin{abstract}

We report the results of a muon-spin rotation ($\mu$SR) experiment to determine the superconducting ground state of the iron-based superconductor CsCa$_2$Fe$_4$As$_4$F$_2$ with $T_{\rm c} \approx 28.3\,$K. This compound is related to the fully-gapped superconductor CaCsFe$_4$As$_4$, but here the Ca-containing spacer layer is replaced with one containing Ca$_2$F$_2$. The temperature evolution of the penetration depth strongly suggests the presence of line nodes and is best modelled by a system consisting of both an $s$- and a $d$-wave gap. We also find a potentially magnetic phase which appears below $\approx 10\,$K but does not appear to compete with the superconductivity. This compound contains the largest alkali atom in this family of superconductors and our results yield a value for the in-plane penetration depth of $\lambda_{ab}(T=0)=244(3)\,$nm.

\end{abstract}

\maketitle

\emph{Introduction ---} Unlike the cuprate high temperature superconductors \cite{Hashimoto2014}, the iron-based superconductors do not appear to have a universal superconducting gap structure: it has been found to vary between families of compounds, as well as within a single family through doping \cite{Reid2012} or pressure \cite{Guguchia2015}. The first iron arsenide superconductor family is a prime example: LaFeAsO$_{1-x}$F$_x$, with a maximum critical temperature $T_{\rm c}=26\,$K \cite{Kamihara2008}, is thought to have either extended $s$-wave pairing \cite{Kuroki2008} with either a reversal of the sign of the order parameter between Fermi surfaces \cite{Mazin2008} or $d$-wave pairing with line nodes \cite{Kuroki2008,Nakai2008,Grafe2008}. The need for a universal picture of gap structures is vital to understand the mechanism with which Cooper pairs are formed in the iron-based superconductors.

The gap structure has been found to vary between members of a single FeAs family. Ba$_{1-x}$K$_x$Fe$_2$As$_2$, part of the 122 family, was initially found to have two $s$-wave gaps \cite{Evtushinsky2008,Khasanov2009}. Subsequent studies, however, found evidence that KFe$_2$As$_2$, which is in the strong hole doping regime in this family, exhibits $d$-wave nodal superconductivity \cite{Fukazawa2009,Hashimoto2010,Thomale2011,Reid2012} which may arise from antiferromagnetic spin fluctuations near the quantum critical point \cite{Dong2010}. It has also been suggested that KFe$_2$As$_2$ has a Fermi surface-selective multigap structure, resulting in $s$-wave superconductivity with accidental line nodes \cite{Okazaki2012}. The replacement of K with Rb or Cs has been reported to conserve the nodal gap structure \cite{Wang2013,Hong2013,Zhang2015}, although there have also been studies to suggest these compounds are $s+s$-wave \cite{Shermadini2010,Shermadini2012}.

There have been several attempts to account for this variation in gap symmetry amongst the iron arsenides. Ref.~\cite{Hirschfeld2011} suggests that, away from optimal doping, subdominant interactions (for example Coulomb interactions) and pair scattering between electron-like Fermi surface sheets increase and frustrate the isotropic $s\pm$ gap, leading to anisotropy and eventually nodes, as evidenced in the 122 family. It has also been theorized that this frustration in FeAs superconductors may lead to time-reversal symmetry breaking (TRSB) and, as a result, an $s+id$ pairing \cite{Lee2009}. Maiti \emph{et al.} \cite{Maiti2011} propose a mechanism based on spin-fluctuation exchange, similar to Ref.~\cite{Hirschfeld2011}. In Ref.~\cite{Maiti2011}, the variation between family members is explained through the degree of electron or hole doping in the compound: small to moderate dopings have an $s$-wave pairing driven by the interpocket electron-hole interaction, strong electron dopings have a $d$-wave pairing from the attraction between electron pockets, and strong hole dopings have either a $d$-wave pairing from attraction within one of the hole pockets or an $s$-wave pairing from the interaction of hole pockets at (0,0). Another cause of the varying gap symmetries may be the height of the pnictogen atom, which alters the competition or cooperation of the spin fluctuation modes \cite{Kuroki2008}.

No such drastic variation is seen in the 1144 family (Ca$A$Fe$_4$As$_4$, $A=\,$K, Rb, or Cs \cite{Iyo2016}), with evidence for multigap $s$-wave superconductivity and a clear absence of nodes in both experimental and theoretical studies \cite{Mou2016,Iida2017,Biswas2017,Cui2017,Cho2017,Suetin2017}. Interestingly, this family has a maximum $T_{\rm c}=35\,$K for $A=\,$Rb, the alkali atom with intermediate size. This contrasts with 1111-type FeAs superconductors \cite{Kamihara2008,Ren2008,Wang2017a}, as well as other alkali metal arsenide superconductors, such as the chain-based compounds $A_2$Cr$_3$As$_3$ ($A=\,$K, Rb, or Cs), where increasing the ionic radius of $A$ leads to a chemical pressure-induced suppression of $T_{\rm c}$ \cite{Bao2015,Tang2015,Zhang2015}.

\begin{figure*}[ht]
	\includegraphics[width=.9\textwidth]{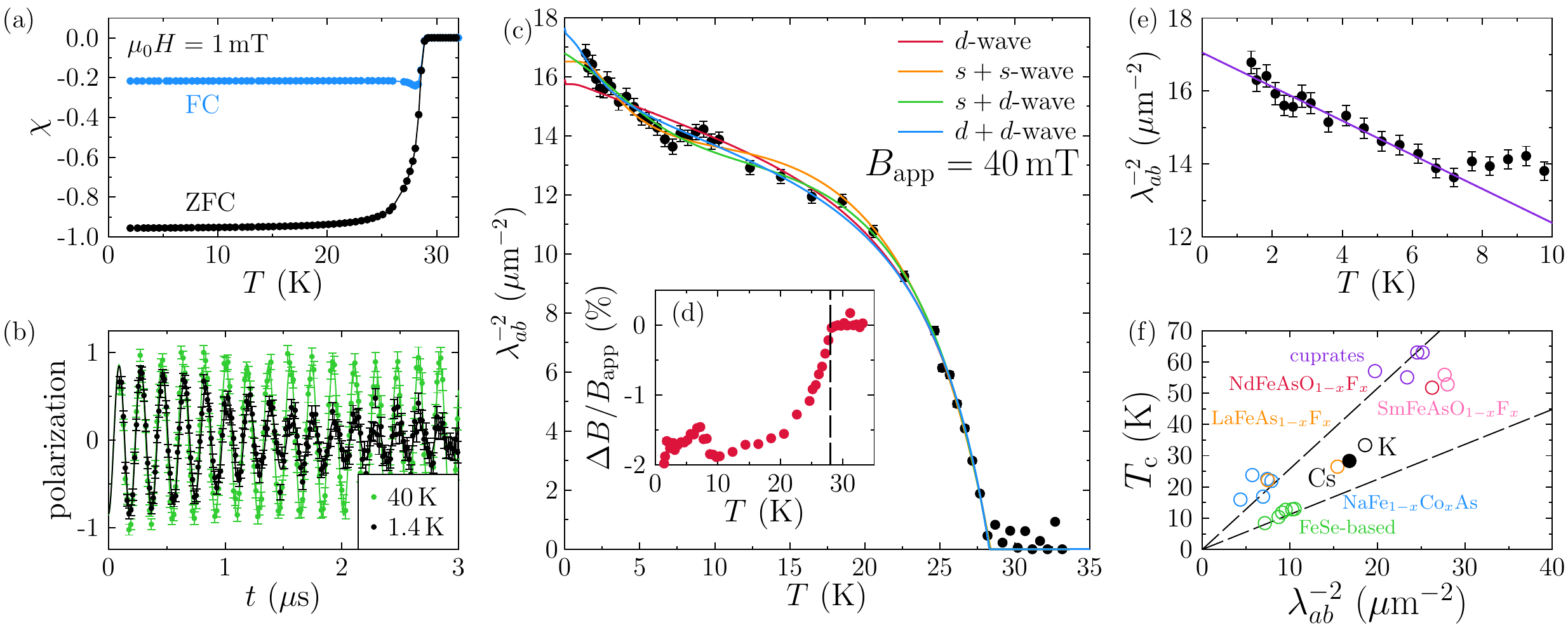}
	\caption{(a) Field cooled (FC) and zero-field cooled (ZFC) magnetic susceptibility of CsCa$_2$Fe$_4$As$_4$F$_2$, $\chi = M/H$, where $M$ is the sample’s magnetization, in a magnetic field of $\mu_0 H = 1\,$mT. (b) Normalized TF-$\mu$SR asymmetry above and below $T_{\rm c}$. Fits using Eq.~\ref{TFeq} are also plotted. (c) Temperature dependence of the inverse square penetration depth in the $a-b$ plane. Fits as in Eq.~\ref{BCSeq} for $d$-, $s+s$-, $s+d$-, and $d+d$-wave superconducting gap structures are also shown. (d) Field shift $\Delta B=B_{\rm app} - B_{\rm SC}$ plotted as a percentage of the applied field $B_{\rm app}$ due to the superconducting vortex lattice in $B_{\rm app} = 40\,$mT. (e) Low-temperature dependence of inverse square penetration depth, with a linear fit to highlight the linear dependence of a gap with line nodes. (f) Uemura plot for various superconductors. The filled black circle indicates CsCa$_2$Fe$_4$As$_4$F$_2$ and the open black circle shows KCa$_2$Fe$_4$As$_4$F$_2$ \cite{Smidman2017} for comparison.}
	\label{TFfig}
\end{figure*}

Another example of a FeAs family with an inverse correlation between $T_{\rm c}$ and alkali metal atomic radius is the 12442 family. The first $A$Ca$_2$Fe$_4$As$_4$F$_2$ compound to be discovered was $A$ = K ($T_{\rm c} = 33\,$K) \cite{Wang2016}, and subsequently $A$ = Rb, Cs ($T_{\rm c} = 31\,$K and $28\,$K respectively) \cite{Wang2017} were also synthesised. This family can be viewed as an intergrowth of $A$Fe$_2$As$_2$ and CaFeAsF layers. Double FeAs layers are sandwiched between $A$ atoms on one side and Ca$_2$F$_2$ on the other, leading to two distinct As sites. These compounds are self-hole-doped, in contrast to other hole-doped FeAs superconductors \cite{Scalapino2012,Jiang2013}. It has been suggested that the complex nature of the electronic structure of these materials could lead to a multiband gap structure \cite{Wang2016a}. $\mu$SR experiments on KCa$_2$Fe$_4$As$_4$F$_2$ have provided further evidence that it is a multigap nodal superconductor with an $s+d$- or $d+d$-wave pairing, with no clear TRSB \cite{Smidman2017}.

In this Rapid Communication, we perform transverse- and zero-field $\mu$SR experiments on CsCa$_2$Fe$_4$As$_4$F$_2$ to determine its superconducting and magnetic properties. Remarkably, we find that the penetration depth of CsCa$_2$Fe$_4$As$_4$F$_2$ does not plateau at low $T$, indicating nodal superconductivity. At intermediate temperatures, there is an inflection point in the temperature dependence of the penetration depth; from this we conclude that CsCa$_2$Fe$_4$As$_4$F$_2$ is a nodal multi-gap superconductor with $s+d$-wave pairing. We also determine that there is no TRSB in this compound, but observe the formation of a potentially magnetic phase below $\approx 10\,$K that does not appear to compete with the superconductivity, although it is possible that it is linked to the opening of the second superconducting gap. Our results suggest a new path to nodal multigap superconductivity in the iron arsenides.

\emph{Experimental Details ---} A polycrystalline sample of CsCa$_2$Fe$_4$As$_4$F$_2$ was synthesised via the solid state reaction in Ref.~\cite{Wang2016}, and was found to have a sharp superconducting transition at 28--29\,K (see Fig.~\ref{TFfig}(a)). The sample in this work was found to have a higher superconducting volume fraction than that in Ref.~\cite{Wang2016}, indicating that this sample has a higher purity.

$\mu$SR experiments \cite{Blundell1999, Yaouanc2011} were performed using a $^3$He cryostat mounted on the MuSR spectrometer at the ISIS pulsed muon facility (Rutherford Appleton Laboratory, UK) \cite{King2013}. Zero-field (ZF) measurements were carried out to check for magnetism, TRSB across the superconducting transition, and to see if an F$\mu$F state (in which the muon forms a hydrogen bond with fluorine atoms in the sample \cite{Brewer1986, Harshman1986}) exists. Transverse-field (TF) measurements, in which an external field (40\,mT) is applied perpendicular to the initial muon polarization, were performed to identify the superconducting ground state. All of the data were analyzed using WiMDA \cite{Pratt2000}.

\emph{Superconductivity ---} Transverse field measurements were performed in an applied field $B_{\rm app}=40\,$mT in order to probe the superconducting ground state. Sample spectra above and below $T_{\rm c}$ are shown in Fig.~\ref{TFfig}(b). There is a clear increase of the relaxation in the superconducting state (compared to the normal state) which arises from the inhomogeneous field distribution of the vortex lattice \cite{Brandt1988}. The data were fitted with the two-component function
\begin{equation} \label{TFeq}
A(t)=A_{B} \cos \left( \gamma_{\mu} B_{\rm app} t + \phi \right) + A_{\rm SC} \cos \left( \gamma_{\mu} B_{\rm SC} t + \phi \right) e^{-\frac{\sigma^2 t^2}{2}},
\end{equation}
where $\gamma_{\mu} = 2\pi \times 135.5\,$MHzT$^{-1}$ is the gyromagnetic ratio of the muon and $\phi$ is a phase related to the detector geometry, with $\phi$ fitted for each of the eight detector groups of the spectrometer. The first term represents those muons which are not implanted into the superconducting volume and precess only in the external magnetic field, and so do not experience any relaxation. This component also includes muons implanted in the sample holder and cryostat. The second term arises from muons in the superconducting volume, which experience a Gaussian broadening $\sigma(T) = \sqrt{\sigma_{\rm SC}^2(T) + \sigma_{\rm nucl}^2}$ consisting of a temperature-dependent component from the vortex lattice, and a temperature-independent component from nuclear moments. By fitting the residual, temperature-independent component of $\sigma$, we find $\sigma_{\rm nucl} = 0.113(3) \mu\rm{s}^{-1}$.
	
The field shifts caused by the vortex lattice $\Delta B~=~B_{\rm SC}~-~B_{\rm app}$ are shown in Fig.~\ref{TFfig}(d). There is a clear negative shift in the peak field as the samples transition into their superconducting states; this is a characteristic feature of the vortex lattice \cite{Brandt1988}. Interestingly, there is a small peak in the field shift at $T \approx 8\,$K, which may be linked to a magnetic phase in the sample that will be discussed below.

In order to extract the penetration depth from $\sigma_{\rm SC}$, a conversion \cite{Brandt2003}
\begin{equation}
\sigma_{\rm SC} = 0.0609 \gamma_{\mu}\phi_0 \lambda_{\rm eff}^{-2}(T),
\end{equation}
can be used, where $\phi_0 = 2.069 \times 10^{-15}$\,Wb is the magnetic flux quantum. This conversion holds for $0.13/\kappa^2 \ll B_{\rm app}/B_{\rm c2} \ll 1$, where $B_{\rm c2}$ is expected to be $> 30\,$T in CsCa$_2$Fe$_4$As$_4$F$_2$ \cite{Wang2017}. As the sample is anisotropic \cite{Wang2017} and polycrystalline, it can therefore be assumed that the effective penetration depth $\lambda_{\rm eff}$ is dominated by the in-plane penetration depth $\lambda_{ab}$ (which is $\ll \lambda_{c}$), where $\lambda_{\rm eff}=3^{1/4}\lambda_{ab}$ \cite{Fesenko1991}. The temperature dependence of $\lambda_{ab}^{-2}$ is plotted in Fig.~\ref{TFfig}(c). The low-$T$ region in Fig.~\ref{TFfig}(e) shows a clear linear dependence of $\lambda_{ab}$, as indicated by the straight line fit. $\lambda_{ab}^{-2}$ for a fully gapped superconductor plateaus at low $T$ as there is not sufficient thermal energy to depopulate the condensate, whereas in nodal superconductors, low-energy excitations are always possible, thereby leading to a linear dependence of $\lambda_{ab}$ at low $T$ \cite{Graf1995}.

The data in Fig.~\ref{TFfig}(c) have been fitted with single- and two-gap BCS models involving $s$- and $d$-wave gaps. The BCS model of the normalized superfluid density of a superconductor is given by \cite{Chandrasekhar1993}:
\begin{equation} \label{BCSeq}
\tilde{n}_{\rm s}(T) = \frac{\lambda^{-2}(T)}{\lambda^{-2}(0)} = 1 + \frac{1}{\pi} \int^{2\pi}_{0} \int^{\infty}_{\Delta(\phi,T)} \frac{\partial f}{\partial E} \frac {E \rm{d}E \rm{d}\phi}{\sqrt{E^2 - \Delta^2 (\phi,T)}},
\end{equation}
where $\Delta(\phi,T)$ is the superconducting gap function, and $f=\left(1+\rm{exp}(E/k_{\rm B}T)\right)^{-1}$ is the Fermi function. The gap function can be approximated as $\Delta(\phi,T) = \Delta(\phi) \tanh \left(1.82 \left[ 1.018 \left(T_{\rm c}/T - 1 \right) \right]^{0.51} \right)$. The angular gap function $\Delta(\phi)=\Delta_0$ for $s$-wave superconductors and $\Delta(\phi) = \Delta_0 \cos (2\phi)$ for $d$-wave (nodal) superconductors. In the case of two gaps, the resulting normalized superfluid density is a weighted sum of the two components: $\tilde{n}_{\rm s}(T) = w\tilde{n}^{\rm (gap~1)}_{\rm s}(T) + (1-w)\tilde{n}^{\rm (gap~2)}_{\rm s}(T)$.

The existence of an inflection point in $\lambda_{ab}(T)$ strongly suggests the existence of two gaps, and is supported by the poor single-gap $d$-wave fit in Fig.~\ref{TFfig}(c) [a single-gap $s$-wave fit did not provide a good fit either, and has not been plotted]. Two-gap fits for $s+s$, $s+d$, and $d+d$ gaps are also shown in Fig.~\ref{TFfig}(c). We find that the fit with the lowest $\chi^2$ is for an $s+d$-wave system with gaps $\Delta_0^{(s)}=7.5(1)\,$meV and $\Delta_0^{(d)}=1.5(1)\,$meV, weighting factor $w=0.73(3)$, $T_{\rm c}=28.31(8)\,$K, and penetration depth $\lambda_{ab} = 244(3)\,$nm. This result has been plotted alongside other common high temperature superconductors in the Uemura plot \cite{Uemura1991} in Fig.~\ref{TFfig}(e); CsCa$_2$Fe$_4$As$_4$F$_2$ falls close to the main scaling line. Both $\lambda_{ab}(0)$ and $T_{\rm c}$ are lower for CsCa$_2$Fe$_4$As$_4$F$_2$ compared to KCa$_2$Fe$_4$As$_4$F$_2$ \cite{Smidman2017}. One key difference between the two compounds is the larger lattice parameter $c$ for CsCa$_2$Fe$_4$As$_4$F$_2$ (32.363(1)\,\AA, compared with 31.007(1)\,\AA\,for KCa$_2$Fe$_4$As$_4$F$_2$ \cite{Wang2016, Wang2017}). However, this geometric difference is insufficient to account for the change in superfluid stiffness (for a contrary case see \cite{Baker2009}). For $A$Ca$_2$Fe$_4$As$_4$F$_2$, a relatively small increase in $c$ ($\approx 5\%$) provides a much larger decrease in $\lambda_{ab}(0)$ ($\approx 20\%$), demonstrating that the increase in $c$ is, at best, only partially responsible for the reduction in $\lambda_{ab}(0)$ and suggesting that other effects associated with the electronic band structure must be contributing more strongly.

\begin{centering}
\begin{figure}[t]
	\includegraphics[width=.5\textwidth]{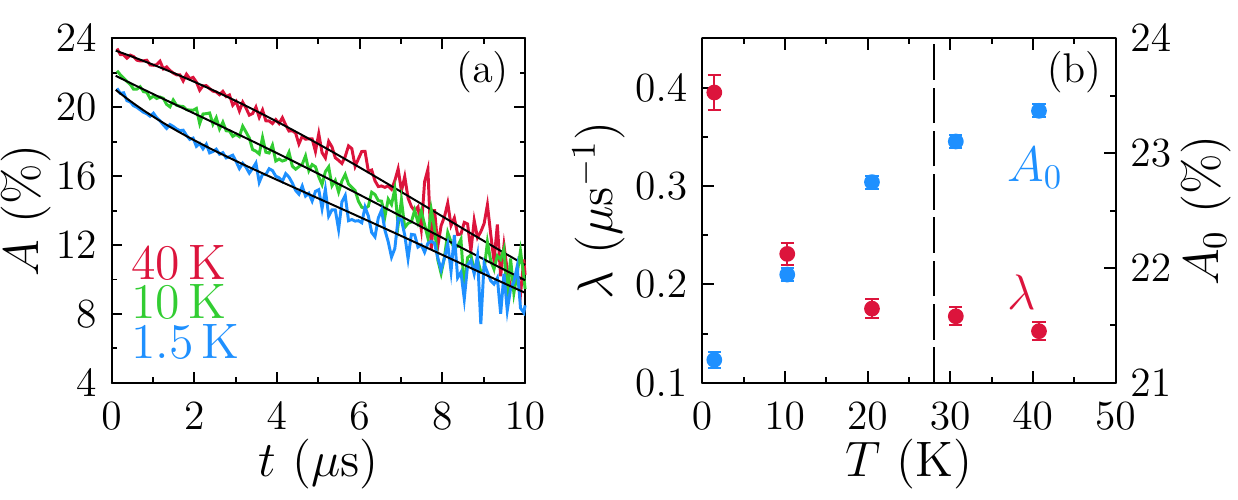}
	\caption{(a) Sample ZF-$\mu$SR data with fits (in black) as in Eq.~\ref{ZFeq}. (b) Temperature dependence of the initial asymmetry $A_0$ and Lorentzian relaxation rate $\lambda$ from fits to the ZF spectra in (a) using Eq.~\ref{ZFeq}. The dashed line marks $T_{\rm c}$.}
	\label{ZFfig}
\end{figure}
\end{centering}

\emph{Magnetism ---} ZF-$\mu$SR measurements were made in order to investigate the possibility of magnetism, TRSB, and an F$\mu$F state. Sample spectra above and below $T_{\rm c}$ are presented in Fig.~\ref{ZFfig}(a). It is clear that there are no oscillations in the spectra, indicating no long range magnetic order or F$\mu$F bonds formed in this compound. The absence of an F$\mu$F signal is common to all fluorine-containing iron-based superconductors studied so far \cite{Luetkens2008, Takeshita2009, Lamura2015}. This may be due either to a muon site close to the FeAs planes (sitting between the As and Ca ions), away from the fluorine ions, or to the metallic nature of the material. The data can be modelled well with
\begin{equation} \label{ZFeq}
A(t) = A_0 \left(a e^{-\sigma^2 t^2 /2} + (1-a) e^{-\lambda t} \right).
\end{equation}
The first component in this function, which accounted for $a \approx 0.76$ of the observed signal, was found to have an approximately constant Gaussian relaxation $\sigma~\approx~0.108(3) \mu $s$^{-1}$, likely arising from fluorine nuclear moments in the sample. The second component contained a temperature-dependent, fast Lorentzian relaxation, plotted in Fig.~\ref{ZFfig}(b). We note a decrease in the initial asymmetry $A_0$, also plotted in Fig.~\ref{ZFfig}(b), which likely arises from an additional fast relaxation.

It is possible that the ZF relaxation may affect the TF spectra in Fig.~\ref{TFfig}(a), similar to the behavior seen in Ref.~\cite{Khasanov2008}. By fitting the data in Fig.~\ref{TFfig}(a) with an additional Lorentzian relaxation, we find a negligible change in the extracted superconducting parameters as well as a larger value of $\chi^2$. It is therefore likely that the source of relaxation in ZF has little effect on the superconductivity observed in the TF data.

There are several possible explanations for the gradual increase of $\lambda$ and decrease of $A_0$ (also plotted in Fig.~\ref{ZFfig}(b)) at low $T$. One possibility is the slowing of fluctuating electronic moments with field width $\Delta$ (where $\lambda = 2 \Delta^2/\nu$ for fluctuation frequency $\nu$). Similar behaviour has been observed in other FeAs superconductors \cite{Khasanov2008}. A source of these moments may be Fe-based impurities, which cannot be ruled out below the 1\% level. The asymmetry of the TF data does not change significantly below $T_{\rm c}$, which is compatible with weak magnetism similar to that observed in heavily doped CaFe$_{1-x}$Co$_x$AsF \cite{Takeshita2009}. We can differentiate true nodal superconductors (where the nodes are imposed by symmetry) to gapped superconductors with `accidental nodes', as impurity scattering acts to broaden symmetry nodes and lifts accidental nodes and therefore removes the residual linear term. If the relaxation arises from Fe-based impurities, this would support the notion that the nodes are symmetry-imposed. Another possible explanation for the behaviour in Fig.~\ref{TFfig}(d) could be due to the opening of the second superconducting gap at approximately the same $T$. This situation may be similar to the $Q$ phase in CeCoIn$_5$ \cite{Agterberg2009, Kato2011} and the pair-density waves in undoped cuprates \cite{Berg2009}, both of which exhibit intertwined magnetic order coupled to $d$-wave superconductivity \cite{Fradkin2015}. The anomalous behaviour in Fig.~\ref{TFfig}(d) could also be explained by such a scenario. Despite the absence of oscillations in this experiment, it is possible that magnetic order does exist with oscillations faster than the resolution of the spectrometer; such behaviour has previously been seen in the near-zero doping of CaFe$_{1-x}$Co$_x$AsF, where oscillations from a magnetic phase with frequency $\approx 25\,$MHz are observed \cite{Takeshita2009}. 

 The signature of a TRSB superconductor is the spontaneous formation of an internal magnetic field below $T_{\rm c}$ \cite{Luke1998}; as this is not seen via a discontinuity in $A_0$ or $\lambda$ at $T_{\rm c}$, we conclude CsCa$_2$Fe$_4$As$_4$F$_2$ does not undergo this transition. We can therefore deduce that the gap symmetry is probably $s+d$-wave, rather than $s+id$.

\emph{Conclusion ---} We have measured the superconducting and magnetic properties of a polycrystalline sample of CsCa$_2$Fe$_4$As$_4$F$_2$. We find that a small potentially magnetic phase exists in the sample, which does not appear to compete with the superconductivity. Remarkably, we find that CsCa$_2$Fe$_4$As$_4$F$_2$ exhibits multigap nodal superconductivity of an $s+d$-wave nature. The ratio $\Delta_0^{(s)}/k_{\rm B}T_{\rm c} \approx 3.1$ of the larger gap suggests strongly coupled superconductivity, although the value of the fitted gap may be sensitive to the area of the nodal region in the Fermi surface \cite{Zhang2011}. Due to the crystallographic structure of this compound, there are two distinct As sites either side of the Fe atoms, and it can be distinguished from the $s$-wave superconductor CaCsFe$_4$As$_4$ by exchanging one of the spacer layers between the FeAs planes. As such, these results indicate a new path from gapped to nodal superconductivity in iron-based superconductors. 

\section{Acknowledgements}

We thank Dr G. Stenning for experimental support. This work is supported by EPSRC grant EP/N023803. F.K.K.K. thanks Lincoln College, Oxford, for a doctoral studentship. D.T.A. thanks the Royal Society of London for UK-China exchange funding. Part of this work was performed at the Science and Technology Facilities Council (STFC) ISIS Facility, Rutherford Appleton Laboratory.

\bibliographystyle{apsrev4-1}
\bibliography{bib1}

\begin{thebibliography}{63}%
\makeatletter
\providecommand \@ifxundefined [1]{%
 \@ifx{#1\undefined}
}%
\providecommand \@ifnum [1]{%
 \ifnum #1\expandafter \@firstoftwo
 \else \expandafter \@secondoftwo
 \fi
}%
\providecommand \@ifx [1]{%
 \ifx #1\expandafter \@firstoftwo
 \else \expandafter \@secondoftwo
 \fi
}%
\providecommand \natexlab [1]{#1}%
\providecommand \enquote  [1]{``#1''}%
\providecommand \bibnamefont  [1]{#1}%
\providecommand \bibfnamefont [1]{#1}%
\providecommand \citenamefont [1]{#1}%
\providecommand \href@noop [0]{\@secondoftwo}%
\providecommand \href [0]{\begingroup \@sanitize@url \@href}%
\providecommand \@href[1]{\@@startlink{#1}\@@href}%
\providecommand \@@href[1]{\endgroup#1\@@endlink}%
\providecommand \@sanitize@url [0]{\catcode `\\12\catcode `\$12\catcode
  `\&12\catcode `\#12\catcode `\^12\catcode `\_12\catcode `\%12\relax}%
\providecommand \@@startlink[1]{}%
\providecommand \@@endlink[0]{}%
\providecommand \url  [0]{\begingroup\@sanitize@url \@url }%
\providecommand \@url [1]{\endgroup\@href {#1}{\urlprefix }}%
\providecommand \urlprefix  [0]{URL }%
\providecommand \Eprint [0]{\href }%
\providecommand \doibase [0]{http://dx.doi.org/}%
\providecommand \selectlanguage [0]{\@gobble}%
\providecommand \bibinfo  [0]{\@secondoftwo}%
\providecommand \bibfield  [0]{\@secondoftwo}%
\providecommand \translation [1]{[#1]}%
\providecommand \BibitemOpen [0]{}%
\providecommand \bibitemStop [0]{}%
\providecommand \bibitemNoStop [0]{.\EOS\space}%
\providecommand \EOS [0]{\spacefactor3000\relax}%
\providecommand \BibitemShut  [1]{\csname bibitem#1\endcsname}%
\let\auto@bib@innerbib\@empty
\bibitem [{\citenamefont {Hashimoto}\ \emph {et~al.}(2014)\citenamefont
  {Hashimoto}, \citenamefont {Vishik}, \citenamefont {He}, \citenamefont
  {Devereaux},\ and\ \citenamefont {Shen}}]{Hashimoto2014}%
  \BibitemOpen
  \bibfield  {author} {\bibinfo {author} {\bibfnamefont {M.}~\bibnamefont
  {Hashimoto}}, \bibinfo {author} {\bibfnamefont {I.~M.}\ \bibnamefont
  {Vishik}}, \bibinfo {author} {\bibfnamefont {R.-H.}\ \bibnamefont {He}},
  \bibinfo {author} {\bibfnamefont {T.~P.}\ \bibnamefont {Devereaux}}, \ and\
  \bibinfo {author} {\bibfnamefont {Z.-X.}\ \bibnamefont {Shen}},\ }\href
  {\doibase 10.1038/nphys3009} {\bibfield  {journal} {\bibinfo  {journal} {Nat.
  Phys.}\ }\textbf {\bibinfo {volume} {10}},\ \bibinfo {pages} {483} (\bibinfo
  {year} {2014})}\BibitemShut {NoStop}%
\bibitem [{\citenamefont {Reid}\ \emph {et~al.}(2012)\citenamefont {Reid},
  \citenamefont {Tanatar}, \citenamefont {Juneau-Fecteau}, \citenamefont
  {Gordon}, \citenamefont {{de Cotret}}, \citenamefont {Doiron-Leyraud},
  \citenamefont {Saito}, \citenamefont {Fukazawa}, \citenamefont {Kohori},
  \citenamefont {Kihou}, \citenamefont {Lee}, \citenamefont {Iyo},
  \citenamefont {Eisaki}, \citenamefont {Prozorov},\ and\ \citenamefont
  {Taillefer}}]{Reid2012}%
  \BibitemOpen
  \bibfield  {author} {\bibinfo {author} {\bibfnamefont {J.~P.}\ \bibnamefont
  {Reid}}, \bibinfo {author} {\bibfnamefont {M.~A.}\ \bibnamefont {Tanatar}},
  \bibinfo {author} {\bibfnamefont {A.}~\bibnamefont {Juneau-Fecteau}},
  \bibinfo {author} {\bibfnamefont {R.~T.}\ \bibnamefont {Gordon}}, \bibinfo
  {author} {\bibfnamefont {S.~R.}\ \bibnamefont {{de Cotret}}}, \bibinfo
  {author} {\bibfnamefont {N.}~\bibnamefont {Doiron-Leyraud}}, \bibinfo
  {author} {\bibfnamefont {T.}~\bibnamefont {Saito}}, \bibinfo {author}
  {\bibfnamefont {H.}~\bibnamefont {Fukazawa}}, \bibinfo {author}
  {\bibfnamefont {Y.}~\bibnamefont {Kohori}}, \bibinfo {author} {\bibfnamefont
  {K.}~\bibnamefont {Kihou}}, \bibinfo {author} {\bibfnamefont {C.~H.}\
  \bibnamefont {Lee}}, \bibinfo {author} {\bibfnamefont {A.}~\bibnamefont
  {Iyo}}, \bibinfo {author} {\bibfnamefont {H.}~\bibnamefont {Eisaki}},
  \bibinfo {author} {\bibfnamefont {R.}~\bibnamefont {Prozorov}}, \ and\
  \bibinfo {author} {\bibfnamefont {L.}~\bibnamefont {Taillefer}},\ }\href
  {https://link.aps.org/doi/10.1103/PhysRevLett.109.087001} {\bibfield
  {journal} {\bibinfo  {journal} {Phys. Rev. Lett.}\ }\textbf {\bibinfo
  {volume} {109}},\ \bibinfo {pages} {087001} (\bibinfo {year}
  {2012})}\BibitemShut {NoStop}%
\bibitem [{\citenamefont {Guguchia}\ \emph {et~al.}(2015)\citenamefont
  {Guguchia}, \citenamefont {Amato}, \citenamefont {Kang}, \citenamefont
  {Luetkens}, \citenamefont {Biswas}, \citenamefont {Prando}, \citenamefont
  {von Rohr}, \citenamefont {Bukowski}, \citenamefont {Shengelaya},
  \citenamefont {Keller}, \citenamefont {Morenzoni}, \citenamefont
  {Fernandes},\ and\ \citenamefont {Khasanov}}]{Guguchia2015}%
  \BibitemOpen
  \bibfield  {author} {\bibinfo {author} {\bibfnamefont {Z.}~\bibnamefont
  {Guguchia}}, \bibinfo {author} {\bibfnamefont {A.}~\bibnamefont {Amato}},
  \bibinfo {author} {\bibfnamefont {J.}~\bibnamefont {Kang}}, \bibinfo {author}
  {\bibfnamefont {H.}~\bibnamefont {Luetkens}}, \bibinfo {author}
  {\bibfnamefont {P.~K.}\ \bibnamefont {Biswas}}, \bibinfo {author}
  {\bibfnamefont {G.}~\bibnamefont {Prando}}, \bibinfo {author} {\bibfnamefont
  {F.}~\bibnamefont {von Rohr}}, \bibinfo {author} {\bibfnamefont
  {Z.}~\bibnamefont {Bukowski}}, \bibinfo {author} {\bibfnamefont
  {A.}~\bibnamefont {Shengelaya}}, \bibinfo {author} {\bibfnamefont
  {H.}~\bibnamefont {Keller}}, \bibinfo {author} {\bibfnamefont
  {E.}~\bibnamefont {Morenzoni}}, \bibinfo {author} {\bibfnamefont
  {R.}~\bibnamefont {Fernandes}}, \ and\ \bibinfo {author} {\bibfnamefont
  {R.}~\bibnamefont {Khasanov}},\ }\href {\doibase 10.1038/ncomms9863}
  {\bibfield  {journal} {\bibinfo  {journal} {Nature Comms.}\ }\textbf
  {\bibinfo {volume} {6}},\ \bibinfo {pages} {8863} (\bibinfo {year}
  {2015})}\BibitemShut {NoStop}%
\bibitem [{\citenamefont {Kamihara}\ \emph {et~al.}(2008)\citenamefont
  {Kamihara}, \citenamefont {Watanabe}, \citenamefont {Hirano},\ and\
  \citenamefont {Hosono}}]{Kamihara2008}%
  \BibitemOpen
  \bibfield  {author} {\bibinfo {author} {\bibfnamefont {Y.}~\bibnamefont
  {Kamihara}}, \bibinfo {author} {\bibfnamefont {T.}~\bibnamefont {Watanabe}},
  \bibinfo {author} {\bibfnamefont {M.}~\bibnamefont {Hirano}}, \ and\ \bibinfo
  {author} {\bibfnamefont {H.}~\bibnamefont {Hosono}},\ }\href {\doibase
  10.1021/ja800073m} {\bibfield  {journal} {\bibinfo  {journal} {J. Am. Chem.
  Soc.}\ }\textbf {\bibinfo {volume} {130}},\ \bibinfo {pages} {3296} (\bibinfo
  {year} {2008})}\BibitemShut {NoStop}%
\bibitem [{\citenamefont {Kuroki}\ \emph {et~al.}(2008)\citenamefont {Kuroki},
  \citenamefont {Onari}, \citenamefont {Arita}, \citenamefont {Usui},
  \citenamefont {Tanaka}, \citenamefont {Kontani},\ and\ \citenamefont
  {Aoki}}]{Kuroki2008}%
  \BibitemOpen
  \bibfield  {author} {\bibinfo {author} {\bibfnamefont {K.}~\bibnamefont
  {Kuroki}}, \bibinfo {author} {\bibfnamefont {S.}~\bibnamefont {Onari}},
  \bibinfo {author} {\bibfnamefont {R.}~\bibnamefont {Arita}}, \bibinfo
  {author} {\bibfnamefont {H.}~\bibnamefont {Usui}}, \bibinfo {author}
  {\bibfnamefont {Y.}~\bibnamefont {Tanaka}}, \bibinfo {author} {\bibfnamefont
  {H.}~\bibnamefont {Kontani}}, \ and\ \bibinfo {author} {\bibfnamefont
  {H.}~\bibnamefont {Aoki}},\ }\href
  {https://journals.aps.org/prl/pdf/10.1103/PhysRevLett.101.087004} {\bibfield
  {journal} {\bibinfo  {journal} {Phys. Rev. Lett.}\ }\textbf {\bibinfo
  {volume} {101}},\ \bibinfo {pages} {087004} (\bibinfo {year}
  {2008})}\BibitemShut {NoStop}%
\bibitem [{\citenamefont {Mazin}\ \emph {et~al.}(2008)\citenamefont {Mazin},
  \citenamefont {Singh}, \citenamefont {Johannes},\ and\ \citenamefont
  {Du}}]{Mazin2008}%
  \BibitemOpen
  \bibfield  {author} {\bibinfo {author} {\bibfnamefont {I.~I.}\ \bibnamefont
  {Mazin}}, \bibinfo {author} {\bibfnamefont {D.~J.}\ \bibnamefont {Singh}},
  \bibinfo {author} {\bibfnamefont {M.~D.}\ \bibnamefont {Johannes}}, \ and\
  \bibinfo {author} {\bibfnamefont {M.~H.}\ \bibnamefont {Du}},\ }\href
  {\doibase 10.1103/PhysRevLett.101.057003} {\bibfield  {journal} {\bibinfo
  {journal} {Phys. Rev. Lett.}\ }\textbf {\bibinfo {volume} {101}},\ \bibinfo
  {pages} {057003} (\bibinfo {year} {2008})}\BibitemShut {NoStop}%
\bibitem [{\citenamefont {Nakai}\ \emph {et~al.}(2008)\citenamefont {Nakai},
  \citenamefont {Ishida}, \citenamefont {Kamihara}, \citenamefont {Hirano},\
  and\ \citenamefont {Hosono}}]{Nakai2008}%
  \BibitemOpen
  \bibfield  {author} {\bibinfo {author} {\bibfnamefont {Y.}~\bibnamefont
  {Nakai}}, \bibinfo {author} {\bibfnamefont {K.}~\bibnamefont {Ishida}},
  \bibinfo {author} {\bibfnamefont {Y.}~\bibnamefont {Kamihara}}, \bibinfo
  {author} {\bibfnamefont {M.}~\bibnamefont {Hirano}}, \ and\ \bibinfo {author}
  {\bibfnamefont {H.}~\bibnamefont {Hosono}},\ }\href {\doibase
  10.1143/JPSJ.77.073701} {\bibfield  {journal} {\bibinfo  {journal} {J. Phys.
  Soc. Jpn.}\ }\textbf {\bibinfo {volume} {77}},\ \bibinfo {pages} {073701}
  (\bibinfo {year} {2008})}\BibitemShut {NoStop}%
\bibitem [{\citenamefont {Grafe}\ \emph {et~al.}(2008)\citenamefont {Grafe},
  \citenamefont {Paar}, \citenamefont {Lang}, \citenamefont {Curro},
  \citenamefont {Behr}, \citenamefont {Werner}, \citenamefont {Hamann-Borrero},
  \citenamefont {Hess}, \citenamefont {Leps}, \citenamefont {Klingeler},\ and\
  \citenamefont {B{\"{u}}chner}}]{Grafe2008}%
  \BibitemOpen
  \bibfield  {author} {\bibinfo {author} {\bibfnamefont {H.~J.}\ \bibnamefont
  {Grafe}}, \bibinfo {author} {\bibfnamefont {D.}~\bibnamefont {Paar}},
  \bibinfo {author} {\bibfnamefont {G.}~\bibnamefont {Lang}}, \bibinfo {author}
  {\bibfnamefont {N.~J.}\ \bibnamefont {Curro}}, \bibinfo {author}
  {\bibfnamefont {G.}~\bibnamefont {Behr}}, \bibinfo {author} {\bibfnamefont
  {J.}~\bibnamefont {Werner}}, \bibinfo {author} {\bibfnamefont
  {J.}~\bibnamefont {Hamann-Borrero}}, \bibinfo {author} {\bibfnamefont
  {C.}~\bibnamefont {Hess}}, \bibinfo {author} {\bibfnamefont {N.}~\bibnamefont
  {Leps}}, \bibinfo {author} {\bibfnamefont {R.}~\bibnamefont {Klingeler}}, \
  and\ \bibinfo {author} {\bibfnamefont {B.}~\bibnamefont {B{\"{u}}chner}},\
  }\href {https://journals.aps.org/prl/pdf/10.1103/PhysRevLett.101.047003}
  {\bibfield  {journal} {\bibinfo  {journal} {Phys. Rev. Lett.}\ }\textbf
  {\bibinfo {volume} {101}},\ \bibinfo {pages} {047003} (\bibinfo {year}
  {2008})}\BibitemShut {NoStop}%
\bibitem [{\citenamefont {Evtushinsky}\ \emph {et~al.}(2009)\citenamefont
  {Evtushinsky}, \citenamefont {Inosov}, \citenamefont {Zabolotnyy},
  \citenamefont {Koitzsch}, \citenamefont {Knupfer}, \citenamefont {Buchner},
  \citenamefont {Viazovska}, \citenamefont {Sun}, \citenamefont {Hinkov},
  \citenamefont {Boris}, \citenamefont {Lin}, \citenamefont {Keimer},
  \citenamefont {Varykhalov}, \citenamefont {Kordyuk},\ and\ \citenamefont
  {Borisenko}}]{Evtushinsky2008}%
  \BibitemOpen
  \bibfield  {author} {\bibinfo {author} {\bibfnamefont {D.~V.}\ \bibnamefont
  {Evtushinsky}}, \bibinfo {author} {\bibfnamefont {D.~S.}\ \bibnamefont
  {Inosov}}, \bibinfo {author} {\bibfnamefont {V.~B.}\ \bibnamefont
  {Zabolotnyy}}, \bibinfo {author} {\bibfnamefont {A.}~\bibnamefont
  {Koitzsch}}, \bibinfo {author} {\bibfnamefont {M.}~\bibnamefont {Knupfer}},
  \bibinfo {author} {\bibfnamefont {B.}~\bibnamefont {Buchner}}, \bibinfo
  {author} {\bibfnamefont {M.~S.}\ \bibnamefont {Viazovska}}, \bibinfo {author}
  {\bibfnamefont {G.~L.}\ \bibnamefont {Sun}}, \bibinfo {author} {\bibfnamefont
  {V.}~\bibnamefont {Hinkov}}, \bibinfo {author} {\bibfnamefont {A.~V.}\
  \bibnamefont {Boris}}, \bibinfo {author} {\bibfnamefont {C.~T.}\ \bibnamefont
  {Lin}}, \bibinfo {author} {\bibfnamefont {B.}~\bibnamefont {Keimer}},
  \bibinfo {author} {\bibfnamefont {A.}~\bibnamefont {Varykhalov}}, \bibinfo
  {author} {\bibfnamefont {A.~A.}\ \bibnamefont {Kordyuk}}, \ and\ \bibinfo
  {author} {\bibfnamefont {S.~V.}\ \bibnamefont {Borisenko}},\ }\href {\doibase
  10.1103/PhysRevB.79.054517} {\bibfield  {journal} {\bibinfo  {journal} {Phys.
  Rev. B}\ }\textbf {\bibinfo {volume} {79}},\ \bibinfo {pages} {054517}
  (\bibinfo {year} {2009})}\BibitemShut {NoStop}%
\bibitem [{\citenamefont {Khasanov}\ \emph {et~al.}(2009)\citenamefont
  {Khasanov}, \citenamefont {Evtushinsky}, \citenamefont {Amato}, \citenamefont
  {Klauss}, \citenamefont {Luetkens}, \citenamefont {Niedermayer},
  \citenamefont {B{\"{u}}chner}, \citenamefont {Sun}, \citenamefont {Lin},
  \citenamefont {Park}, \citenamefont {Inosov},\ and\ \citenamefont
  {Hinkov}}]{Khasanov2009}%
  \BibitemOpen
  \bibfield  {author} {\bibinfo {author} {\bibfnamefont {R.}~\bibnamefont
  {Khasanov}}, \bibinfo {author} {\bibfnamefont {D.~V.}\ \bibnamefont
  {Evtushinsky}}, \bibinfo {author} {\bibfnamefont {A.}~\bibnamefont {Amato}},
  \bibinfo {author} {\bibfnamefont {H.~H.}\ \bibnamefont {Klauss}}, \bibinfo
  {author} {\bibfnamefont {H.}~\bibnamefont {Luetkens}}, \bibinfo {author}
  {\bibfnamefont {C.}~\bibnamefont {Niedermayer}}, \bibinfo {author}
  {\bibfnamefont {B.}~\bibnamefont {B{\"{u}}chner}}, \bibinfo {author}
  {\bibfnamefont {G.~L.}\ \bibnamefont {Sun}}, \bibinfo {author} {\bibfnamefont
  {C.~T.}\ \bibnamefont {Lin}}, \bibinfo {author} {\bibfnamefont {J.~T.}\
  \bibnamefont {Park}}, \bibinfo {author} {\bibfnamefont {D.~S.}\ \bibnamefont
  {Inosov}}, \ and\ \bibinfo {author} {\bibfnamefont {V.}~\bibnamefont
  {Hinkov}},\ }\href
  {https://journals.aps.org/prl/pdf/10.1103/PhysRevLett.102.187005} {\bibfield
  {journal} {\bibinfo  {journal} {Phys. Rev. Lett.}\ }\textbf {\bibinfo
  {volume} {102}},\ \bibinfo {pages} {187005} (\bibinfo {year}
  {2009})}\BibitemShut {NoStop}%
\bibitem [{\citenamefont {Fukazawa}\ \emph {et~al.}(2009)\citenamefont
  {Fukazawa}, \citenamefont {Yamada}, \citenamefont {Kondo}, \citenamefont
  {Saito}, \citenamefont {Kohori}, \citenamefont {Kuga}, \citenamefont
  {Matsumoto}, \citenamefont {Nakatsuji}, \citenamefont {Kito}, \citenamefont
  {Shirage}, \citenamefont {Kihou}, \citenamefont {Takeshita}, \citenamefont
  {Lee}, \citenamefont {Iyo},\ and\ \citenamefont {Eisaki}}]{Fukazawa2009}%
  \BibitemOpen
  \bibfield  {author} {\bibinfo {author} {\bibfnamefont {H.}~\bibnamefont
  {Fukazawa}}, \bibinfo {author} {\bibfnamefont {Y.}~\bibnamefont {Yamada}},
  \bibinfo {author} {\bibfnamefont {K.}~\bibnamefont {Kondo}}, \bibinfo
  {author} {\bibfnamefont {T.}~\bibnamefont {Saito}}, \bibinfo {author}
  {\bibfnamefont {Y.}~\bibnamefont {Kohori}}, \bibinfo {author} {\bibfnamefont
  {K.}~\bibnamefont {Kuga}}, \bibinfo {author} {\bibfnamefont {Y.}~\bibnamefont
  {Matsumoto}}, \bibinfo {author} {\bibfnamefont {S.}~\bibnamefont
  {Nakatsuji}}, \bibinfo {author} {\bibfnamefont {H.}~\bibnamefont {Kito}},
  \bibinfo {author} {\bibfnamefont {P.~M.}\ \bibnamefont {Shirage}}, \bibinfo
  {author} {\bibfnamefont {K.}~\bibnamefont {Kihou}}, \bibinfo {author}
  {\bibfnamefont {N.}~\bibnamefont {Takeshita}}, \bibinfo {author}
  {\bibfnamefont {C.~H.}\ \bibnamefont {Lee}}, \bibinfo {author} {\bibfnamefont
  {A.}~\bibnamefont {Iyo}}, \ and\ \bibinfo {author} {\bibfnamefont
  {H.}~\bibnamefont {Eisaki}},\ }\href {\doibase 10.1143/JPSJ.78.083712}
  {\bibfield  {journal} {\bibinfo  {journal} {J. Phys. Soc. Jpn.}\ }\textbf
  {\bibinfo {volume} {78}},\ \bibinfo {pages} {083712} (\bibinfo {year}
  {2009})}\BibitemShut {NoStop}%
\bibitem [{\citenamefont {Hashimoto}\ \emph {et~al.}(2010)\citenamefont
  {Hashimoto}, \citenamefont {Serafin}, \citenamefont {Tonegawa}, \citenamefont
  {Katsumata}, \citenamefont {Okazaki}, \citenamefont {Saito}, \citenamefont
  {Fukazawa}, \citenamefont {Kohori}, \citenamefont {Kihou}, \citenamefont
  {Lee}, \citenamefont {Iyo}, \citenamefont {Eisaki}, \citenamefont {Ikeda},
  \citenamefont {Matsuda}, \citenamefont {Carrington},\ and\ \citenamefont
  {Shibauchi}}]{Hashimoto2010}%
  \BibitemOpen
  \bibfield  {author} {\bibinfo {author} {\bibfnamefont {K.}~\bibnamefont
  {Hashimoto}}, \bibinfo {author} {\bibfnamefont {A.}~\bibnamefont {Serafin}},
  \bibinfo {author} {\bibfnamefont {S.}~\bibnamefont {Tonegawa}}, \bibinfo
  {author} {\bibfnamefont {R.}~\bibnamefont {Katsumata}}, \bibinfo {author}
  {\bibfnamefont {R.}~\bibnamefont {Okazaki}}, \bibinfo {author} {\bibfnamefont
  {T.}~\bibnamefont {Saito}}, \bibinfo {author} {\bibfnamefont
  {H.}~\bibnamefont {Fukazawa}}, \bibinfo {author} {\bibfnamefont
  {Y.}~\bibnamefont {Kohori}}, \bibinfo {author} {\bibfnamefont
  {K.}~\bibnamefont {Kihou}}, \bibinfo {author} {\bibfnamefont {C.~H.}\
  \bibnamefont {Lee}}, \bibinfo {author} {\bibfnamefont {A.}~\bibnamefont
  {Iyo}}, \bibinfo {author} {\bibfnamefont {H.}~\bibnamefont {Eisaki}},
  \bibinfo {author} {\bibfnamefont {H.}~\bibnamefont {Ikeda}}, \bibinfo
  {author} {\bibfnamefont {Y.}~\bibnamefont {Matsuda}}, \bibinfo {author}
  {\bibfnamefont {A.}~\bibnamefont {Carrington}}, \ and\ \bibinfo {author}
  {\bibfnamefont {T.}~\bibnamefont {Shibauchi}},\ }\href {\doibase
  10.1103/PhysRevB.82.014526} {\bibfield  {journal} {\bibinfo  {journal} {Phys.
  Rev. B}\ }\textbf {\bibinfo {volume} {82}},\ \bibinfo {pages} {014526}
  (\bibinfo {year} {2010})}\BibitemShut {NoStop}%
\bibitem [{\citenamefont {Thomale}\ \emph {et~al.}(2011)\citenamefont
  {Thomale}, \citenamefont {Platt}, \citenamefont {Hanke}, \citenamefont {Hu},\
  and\ \citenamefont {Bernevig}}]{Thomale2011}%
  \BibitemOpen
  \bibfield  {author} {\bibinfo {author} {\bibfnamefont {R.}~\bibnamefont
  {Thomale}}, \bibinfo {author} {\bibfnamefont {C.}~\bibnamefont {Platt}},
  \bibinfo {author} {\bibfnamefont {W.}~\bibnamefont {Hanke}}, \bibinfo
  {author} {\bibfnamefont {J.}~\bibnamefont {Hu}}, \ and\ \bibinfo {author}
  {\bibfnamefont {B.~A.}\ \bibnamefont {Bernevig}},\ }\href {\doibase
  10.1103/PhysRevLett.107.117001} {\bibfield  {journal} {\bibinfo  {journal}
  {Phys. Rev. Lett.}\ }\textbf {\bibinfo {volume} {107}},\ \bibinfo {pages}
  {117001} (\bibinfo {year} {2011})}\BibitemShut {NoStop}%
\bibitem [{\citenamefont {Dong}\ \emph {et~al.}(2010)\citenamefont {Dong},
  \citenamefont {Zhou}, \citenamefont {Guan}, \citenamefont {Zhang},
  \citenamefont {Dai}, \citenamefont {Qiu}, \citenamefont {Wang}, \citenamefont
  {He}, \citenamefont {Chen},\ and\ \citenamefont {Li}}]{Dong2010}%
  \BibitemOpen
  \bibfield  {author} {\bibinfo {author} {\bibfnamefont {J.~K.}\ \bibnamefont
  {Dong}}, \bibinfo {author} {\bibfnamefont {S.~Y.}\ \bibnamefont {Zhou}},
  \bibinfo {author} {\bibfnamefont {T.~Y.}\ \bibnamefont {Guan}}, \bibinfo
  {author} {\bibfnamefont {H.}~\bibnamefont {Zhang}}, \bibinfo {author}
  {\bibfnamefont {Y.~F.}\ \bibnamefont {Dai}}, \bibinfo {author} {\bibfnamefont
  {X.}~\bibnamefont {Qiu}}, \bibinfo {author} {\bibfnamefont {X.~F.}\
  \bibnamefont {Wang}}, \bibinfo {author} {\bibfnamefont {Y.}~\bibnamefont
  {He}}, \bibinfo {author} {\bibfnamefont {X.~H.}\ \bibnamefont {Chen}}, \ and\
  \bibinfo {author} {\bibfnamefont {S.~Y.}\ \bibnamefont {Li}},\ }\href
  {\doibase 10.1103/PhysRevLett.104.087005} {\bibfield  {journal} {\bibinfo
  {journal} {Phys. Rev. Lett.}\ }\textbf {\bibinfo {volume} {104}},\ \bibinfo
  {pages} {087005} (\bibinfo {year} {2010})}\BibitemShut {NoStop}%
\bibitem [{\citenamefont {Okazaki}\ \emph {et~al.}(2012)\citenamefont
  {Okazaki}, \citenamefont {Ota}, \citenamefont {Kotani}, \citenamefont
  {Malaeb}, \citenamefont {Ishida}, \citenamefont {Shimojima}, \citenamefont
  {Kiss}, \citenamefont {Watanabe}, \citenamefont {Chen}, \citenamefont
  {Kihou}, \citenamefont {Lee}, \citenamefont {Iyo}, \citenamefont {Eisaki},
  \citenamefont {Saito}, \citenamefont {Fukazawa}, \citenamefont {Kohori},
  \citenamefont {Hashimoto}, \citenamefont {Shibauchi}, \citenamefont
  {Matsuda}, \citenamefont {Ikeda}, \citenamefont {Miyahara}, \citenamefont
  {Arita}, \citenamefont {Chainani},\ and\ \citenamefont {Shin}}]{Okazaki2012}%
  \BibitemOpen
  \bibfield  {author} {\bibinfo {author} {\bibfnamefont {K.}~\bibnamefont
  {Okazaki}}, \bibinfo {author} {\bibfnamefont {Y.}~\bibnamefont {Ota}},
  \bibinfo {author} {\bibfnamefont {Y.}~\bibnamefont {Kotani}}, \bibinfo
  {author} {\bibfnamefont {W.}~\bibnamefont {Malaeb}}, \bibinfo {author}
  {\bibfnamefont {Y.}~\bibnamefont {Ishida}}, \bibinfo {author} {\bibfnamefont
  {T.}~\bibnamefont {Shimojima}}, \bibinfo {author} {\bibfnamefont
  {T.}~\bibnamefont {Kiss}}, \bibinfo {author} {\bibfnamefont {S.}~\bibnamefont
  {Watanabe}}, \bibinfo {author} {\bibfnamefont {C.-T.}\ \bibnamefont {Chen}},
  \bibinfo {author} {\bibfnamefont {K.}~\bibnamefont {Kihou}}, \bibinfo
  {author} {\bibfnamefont {C.~H.}\ \bibnamefont {Lee}}, \bibinfo {author}
  {\bibfnamefont {A.}~\bibnamefont {Iyo}}, \bibinfo {author} {\bibfnamefont
  {H.}~\bibnamefont {Eisaki}}, \bibinfo {author} {\bibfnamefont
  {T.}~\bibnamefont {Saito}}, \bibinfo {author} {\bibfnamefont
  {H.}~\bibnamefont {Fukazawa}}, \bibinfo {author} {\bibfnamefont
  {Y.}~\bibnamefont {Kohori}}, \bibinfo {author} {\bibfnamefont
  {K.}~\bibnamefont {Hashimoto}}, \bibinfo {author} {\bibfnamefont
  {T.}~\bibnamefont {Shibauchi}}, \bibinfo {author} {\bibfnamefont
  {Y.}~\bibnamefont {Matsuda}}, \bibinfo {author} {\bibfnamefont
  {H.}~\bibnamefont {Ikeda}}, \bibinfo {author} {\bibfnamefont
  {H.}~\bibnamefont {Miyahara}}, \bibinfo {author} {\bibfnamefont
  {R.}~\bibnamefont {Arita}}, \bibinfo {author} {\bibfnamefont
  {A.}~\bibnamefont {Chainani}}, \ and\ \bibinfo {author} {\bibfnamefont
  {S.}~\bibnamefont {Shin}},\ }\href {\doibase 10.1126/science.1222793}
  {\bibfield  {journal} {\bibinfo  {journal} {Science}\ }\textbf {\bibinfo
  {volume} {337}},\ \bibinfo {pages} {1314} (\bibinfo {year}
  {2012})}\BibitemShut {NoStop}%
\bibitem [{\citenamefont {Wang}\ \emph {et~al.}(2013)\citenamefont {Wang},
  \citenamefont {Pan}, \citenamefont {Luo}, \citenamefont {Chen}, \citenamefont
  {Yan}, \citenamefont {Ying}, \citenamefont {Ye}, \citenamefont {Cheng},
  \citenamefont {Hong}, \citenamefont {Li},\ and\ \citenamefont
  {Chen}}]{Wang2013}%
  \BibitemOpen
  \bibfield  {author} {\bibinfo {author} {\bibfnamefont {A.~F.}\ \bibnamefont
  {Wang}}, \bibinfo {author} {\bibfnamefont {B.~Y.}\ \bibnamefont {Pan}},
  \bibinfo {author} {\bibfnamefont {X.~G.}\ \bibnamefont {Luo}}, \bibinfo
  {author} {\bibfnamefont {F.}~\bibnamefont {Chen}}, \bibinfo {author}
  {\bibfnamefont {Y.~J.}\ \bibnamefont {Yan}}, \bibinfo {author} {\bibfnamefont
  {J.~J.}\ \bibnamefont {Ying}}, \bibinfo {author} {\bibfnamefont {G.~J.}\
  \bibnamefont {Ye}}, \bibinfo {author} {\bibfnamefont {P.}~\bibnamefont
  {Cheng}}, \bibinfo {author} {\bibfnamefont {X.~C.}\ \bibnamefont {Hong}},
  \bibinfo {author} {\bibfnamefont {S.~Y.}\ \bibnamefont {Li}}, \ and\ \bibinfo
  {author} {\bibfnamefont {X.~H.}\ \bibnamefont {Chen}},\ }\href {\doibase
  10.1103/PhysRevB.87.214509} {\bibfield  {journal} {\bibinfo  {journal} {Phys.
  Rev. B}\ }\textbf {\bibinfo {volume} {87}},\ \bibinfo {pages} {214509}
  (\bibinfo {year} {2013})}\BibitemShut {NoStop}%
\bibitem [{\citenamefont {Hong}\ \emph {et~al.}(2013)\citenamefont {Hong},
  \citenamefont {Li}, \citenamefont {Pan}, \citenamefont {He}, \citenamefont
  {Wang}, \citenamefont {Luo}, \citenamefont {Chen},\ and\ \citenamefont
  {Li}}]{Hong2013}%
  \BibitemOpen
  \bibfield  {author} {\bibinfo {author} {\bibfnamefont {X.~C.}\ \bibnamefont
  {Hong}}, \bibinfo {author} {\bibfnamefont {X.~L.}\ \bibnamefont {Li}},
  \bibinfo {author} {\bibfnamefont {B.~Y.}\ \bibnamefont {Pan}}, \bibinfo
  {author} {\bibfnamefont {L.~P.}\ \bibnamefont {He}}, \bibinfo {author}
  {\bibfnamefont {A.~F.}\ \bibnamefont {Wang}}, \bibinfo {author}
  {\bibfnamefont {X.~G.}\ \bibnamefont {Luo}}, \bibinfo {author} {\bibfnamefont
  {X.~H.}\ \bibnamefont {Chen}}, \ and\ \bibinfo {author} {\bibfnamefont
  {S.~Y.}\ \bibnamefont {Li}},\ }\href {\doibase 10.1103/PhysRevB.87.144502}
  {\bibfield  {journal} {\bibinfo  {journal} {Phys. Rev. B}\ }\textbf {\bibinfo
  {volume} {87}},\ \bibinfo {pages} {144502} (\bibinfo {year}
  {2013})}\BibitemShut {NoStop}%
\bibitem [{\citenamefont {Zhang}\ \emph {et~al.}(2015)\citenamefont {Zhang},
  \citenamefont {Wang}, \citenamefont {Hong}, \citenamefont {Zhang},
  \citenamefont {Pan}, \citenamefont {Pan}, \citenamefont {Xu}, \citenamefont
  {Luo}, \citenamefont {Chen},\ and\ \citenamefont {Li}}]{Zhang2015}%
  \BibitemOpen
  \bibfield  {author} {\bibinfo {author} {\bibfnamefont {Z.}~\bibnamefont
  {Zhang}}, \bibinfo {author} {\bibfnamefont {A.~F.}\ \bibnamefont {Wang}},
  \bibinfo {author} {\bibfnamefont {X.~C.}\ \bibnamefont {Hong}}, \bibinfo
  {author} {\bibfnamefont {J.}~\bibnamefont {Zhang}}, \bibinfo {author}
  {\bibfnamefont {B.~Y.}\ \bibnamefont {Pan}}, \bibinfo {author} {\bibfnamefont
  {J.}~\bibnamefont {Pan}}, \bibinfo {author} {\bibfnamefont {Y.}~\bibnamefont
  {Xu}}, \bibinfo {author} {\bibfnamefont {X.~G.}\ \bibnamefont {Luo}},
  \bibinfo {author} {\bibfnamefont {X.~H.}\ \bibnamefont {Chen}}, \ and\
  \bibinfo {author} {\bibfnamefont {S.~Y.}\ \bibnamefont {Li}},\ }\href
  {\doibase 10.1103/PhysRevB.91.024502} {\bibfield  {journal} {\bibinfo
  {journal} {Phys. Rev. B}\ }\textbf {\bibinfo {volume} {91}},\ \bibinfo
  {pages} {024502} (\bibinfo {year} {2015})}\BibitemShut {NoStop}%
\bibitem [{\citenamefont {Shermadini}\ \emph {et~al.}(2010)\citenamefont
  {Shermadini}, \citenamefont {Kanter}, \citenamefont {Baines}, \citenamefont
  {Bendele}, \citenamefont {Bukowski}, \citenamefont {Khasanov}, \citenamefont
  {Klauss}, \citenamefont {Luetkens}, \citenamefont {Maeter}, \citenamefont
  {Pascua}, \citenamefont {Batlogg},\ and\ \citenamefont
  {Amato}}]{Shermadini2010}%
  \BibitemOpen
  \bibfield  {author} {\bibinfo {author} {\bibfnamefont {Z.}~\bibnamefont
  {Shermadini}}, \bibinfo {author} {\bibfnamefont {J.}~\bibnamefont {Kanter}},
  \bibinfo {author} {\bibfnamefont {C.}~\bibnamefont {Baines}}, \bibinfo
  {author} {\bibfnamefont {M.}~\bibnamefont {Bendele}}, \bibinfo {author}
  {\bibfnamefont {Z.}~\bibnamefont {Bukowski}}, \bibinfo {author}
  {\bibfnamefont {R.}~\bibnamefont {Khasanov}}, \bibinfo {author}
  {\bibfnamefont {H.-H.}\ \bibnamefont {Klauss}}, \bibinfo {author}
  {\bibfnamefont {H.}~\bibnamefont {Luetkens}}, \bibinfo {author}
  {\bibfnamefont {H.}~\bibnamefont {Maeter}}, \bibinfo {author} {\bibfnamefont
  {G.}~\bibnamefont {Pascua}}, \bibinfo {author} {\bibfnamefont
  {B.}~\bibnamefont {Batlogg}}, \ and\ \bibinfo {author} {\bibfnamefont
  {A.}~\bibnamefont {Amato}},\ }\href {\doibase 10.1103/PhysRevB.82.144527}
  {\bibfield  {journal} {\bibinfo  {journal} {Phys. Rev. B}\ }\textbf {\bibinfo
  {volume} {82}},\ \bibinfo {pages} {144527} (\bibinfo {year}
  {2010})}\BibitemShut {NoStop}%
\bibitem [{\citenamefont {Shermadini}\ \emph {et~al.}(2012)\citenamefont
  {Shermadini}, \citenamefont {Luetkens}, \citenamefont {Maisuradze},
  \citenamefont {Khasanov}, \citenamefont {Bukowski}, \citenamefont {Klauss},\
  and\ \citenamefont {Amato}}]{Shermadini2012}%
  \BibitemOpen
  \bibfield  {author} {\bibinfo {author} {\bibfnamefont {Z.}~\bibnamefont
  {Shermadini}}, \bibinfo {author} {\bibfnamefont {H.}~\bibnamefont
  {Luetkens}}, \bibinfo {author} {\bibfnamefont {A.}~\bibnamefont
  {Maisuradze}}, \bibinfo {author} {\bibfnamefont {R.}~\bibnamefont
  {Khasanov}}, \bibinfo {author} {\bibfnamefont {Z.}~\bibnamefont {Bukowski}},
  \bibinfo {author} {\bibfnamefont {H.~H.}\ \bibnamefont {Klauss}}, \ and\
  \bibinfo {author} {\bibfnamefont {A.}~\bibnamefont {Amato}},\ }\href
  {\doibase 10.1103/PhysRevB.86.174516} {\bibfield  {journal} {\bibinfo
  {journal} {Phys. Rev. B}\ }\textbf {\bibinfo {volume} {86}},\ \bibinfo
  {pages} {174516} (\bibinfo {year} {2012})}\BibitemShut {NoStop}%
\bibitem [{\citenamefont {Hirschfeld}\ \emph {et~al.}(2011)\citenamefont
  {Hirschfeld}, \citenamefont {Korshunov},\ and\ \citenamefont
  {Mazin}}]{Hirschfeld2011}%
  \BibitemOpen
  \bibfield  {author} {\bibinfo {author} {\bibfnamefont {P.~J.}\ \bibnamefont
  {Hirschfeld}}, \bibinfo {author} {\bibfnamefont {M.~M.}\ \bibnamefont
  {Korshunov}}, \ and\ \bibinfo {author} {\bibfnamefont {I.~I.}\ \bibnamefont
  {Mazin}},\ }\href {\doibase 10.1088/0034-4885/74/12/124508} {\bibfield
  {journal} {\bibinfo  {journal} {Rep. Prog. Phys.}\ }\textbf {\bibinfo
  {volume} {74}},\ \bibinfo {pages} {124508} (\bibinfo {year}
  {2011})}\BibitemShut {NoStop}%
\bibitem [{\citenamefont {Lee}\ \emph {et~al.}(2009)\citenamefont {Lee},
  \citenamefont {Zhang},\ and\ \citenamefont {Wu}}]{Lee2009}%
  \BibitemOpen
  \bibfield  {author} {\bibinfo {author} {\bibfnamefont {W.~C.}\ \bibnamefont
  {Lee}}, \bibinfo {author} {\bibfnamefont {S.~C.}\ \bibnamefont {Zhang}}, \
  and\ \bibinfo {author} {\bibfnamefont {C.}~\bibnamefont {Wu}},\ }\href
  {\doibase 10.1103/PhysRevLett.102.217002} {\bibfield  {journal} {\bibinfo
  {journal} {Phys. Rev. Lett.}\ }\textbf {\bibinfo {volume} {102}},\ \bibinfo
  {pages} {217002} (\bibinfo {year} {2009})}\BibitemShut {NoStop}%
\bibitem [{\citenamefont {Maiti}\ \emph {et~al.}(2011)\citenamefont {Maiti},
  \citenamefont {Korshunov}, \citenamefont {Maier}, \citenamefont
  {Hirschfeld},\ and\ \citenamefont {Chubukov}}]{Maiti2011}%
  \BibitemOpen
  \bibfield  {author} {\bibinfo {author} {\bibfnamefont {S.}~\bibnamefont
  {Maiti}}, \bibinfo {author} {\bibfnamefont {M.~M.}\ \bibnamefont
  {Korshunov}}, \bibinfo {author} {\bibfnamefont {T.~A.}\ \bibnamefont
  {Maier}}, \bibinfo {author} {\bibfnamefont {P.~J.}\ \bibnamefont
  {Hirschfeld}}, \ and\ \bibinfo {author} {\bibfnamefont {A.~V.}\ \bibnamefont
  {Chubukov}},\ }\href {\doibase 10.1103/PhysRevLett.107.147002} {\bibfield
  {journal} {\bibinfo  {journal} {Phys. Rev. Lett.}\ }\textbf {\bibinfo
  {volume} {107}},\ \bibinfo {pages} {147002} (\bibinfo {year}
  {2011})}\BibitemShut {NoStop}%
\bibitem [{\citenamefont {Iyo}\ \emph {et~al.}(2016)\citenamefont {Iyo},
  \citenamefont {Kawashima}, \citenamefont {Kinjo}, \citenamefont {Nishio},
  \citenamefont {Ishida}, \citenamefont {Fujihisa}, \citenamefont {Gotoh},
  \citenamefont {Kihou}, \citenamefont {Eisaki},\ and\ \citenamefont
  {Yoshida}}]{Iyo2016}%
  \BibitemOpen
  \bibfield  {author} {\bibinfo {author} {\bibfnamefont {A.}~\bibnamefont
  {Iyo}}, \bibinfo {author} {\bibfnamefont {K.}~\bibnamefont {Kawashima}},
  \bibinfo {author} {\bibfnamefont {T.}~\bibnamefont {Kinjo}}, \bibinfo
  {author} {\bibfnamefont {T.}~\bibnamefont {Nishio}}, \bibinfo {author}
  {\bibfnamefont {S.}~\bibnamefont {Ishida}}, \bibinfo {author} {\bibfnamefont
  {H.}~\bibnamefont {Fujihisa}}, \bibinfo {author} {\bibfnamefont
  {Y.}~\bibnamefont {Gotoh}}, \bibinfo {author} {\bibfnamefont
  {K.}~\bibnamefont {Kihou}}, \bibinfo {author} {\bibfnamefont
  {H.}~\bibnamefont {Eisaki}}, \ and\ \bibinfo {author} {\bibfnamefont
  {Y.}~\bibnamefont {Yoshida}},\ }\href {\doibase 10.1021/jacs.5b12571}
  {\bibfield  {journal} {\bibinfo  {journal} {J. Am. Chem. Soc.}\ }\textbf
  {\bibinfo {volume} {138}},\ \bibinfo {pages} {3410} (\bibinfo {year}
  {2016})}\BibitemShut {NoStop}%
\bibitem [{\citenamefont {Mou}\ \emph {et~al.}(2016)\citenamefont {Mou},
  \citenamefont {Kong}, \citenamefont {Meier}, \citenamefont {Lochner},
  \citenamefont {Wang}, \citenamefont {Lin}, \citenamefont {Wu}, \citenamefont
  {Budko}, \citenamefont {Eremin}, \citenamefont {Johnson}, \citenamefont
  {Canfield},\ and\ \citenamefont {Kaminski}}]{Mou2016}%
  \BibitemOpen
  \bibfield  {author} {\bibinfo {author} {\bibfnamefont {D.}~\bibnamefont
  {Mou}}, \bibinfo {author} {\bibfnamefont {T.}~\bibnamefont {Kong}}, \bibinfo
  {author} {\bibfnamefont {W.~R.}\ \bibnamefont {Meier}}, \bibinfo {author}
  {\bibfnamefont {F.}~\bibnamefont {Lochner}}, \bibinfo {author} {\bibfnamefont
  {L.~L.}\ \bibnamefont {Wang}}, \bibinfo {author} {\bibfnamefont
  {Q.}~\bibnamefont {Lin}}, \bibinfo {author} {\bibfnamefont {Y.}~\bibnamefont
  {Wu}}, \bibinfo {author} {\bibfnamefont {S.~L.}\ \bibnamefont {Budko}},
  \bibinfo {author} {\bibfnamefont {I.}~\bibnamefont {Eremin}}, \bibinfo
  {author} {\bibfnamefont {D.~D.}\ \bibnamefont {Johnson}}, \bibinfo {author}
  {\bibfnamefont {P.~C.}\ \bibnamefont {Canfield}}, \ and\ \bibinfo {author}
  {\bibfnamefont {A.}~\bibnamefont {Kaminski}},\ }\href {\doibase
  10.1103/PhysRevLett.117.277001} {\bibfield  {journal} {\bibinfo  {journal}
  {Phys. Rev. Lett.}\ }\textbf {\bibinfo {volume} {117}},\ \bibinfo {pages}
  {277001} (\bibinfo {year} {2016})}\BibitemShut {NoStop}%
\bibitem [{\citenamefont {Iida}\ \emph {et~al.}(2017)\citenamefont {Iida},
  \citenamefont {Ishikado}, \citenamefont {Nagai}, \citenamefont {Yoshida},
  \citenamefont {Christianson}, \citenamefont {Murai}, \citenamefont
  {Kawashima}, \citenamefont {Yoshida}, \citenamefont {Eisaki},\ and\
  \citenamefont {Iyo}}]{Iida2017}%
  \BibitemOpen
  \bibfield  {author} {\bibinfo {author} {\bibfnamefont {K.}~\bibnamefont
  {Iida}}, \bibinfo {author} {\bibfnamefont {M.}~\bibnamefont {Ishikado}},
  \bibinfo {author} {\bibfnamefont {Y.}~\bibnamefont {Nagai}}, \bibinfo
  {author} {\bibfnamefont {H.}~\bibnamefont {Yoshida}}, \bibinfo {author}
  {\bibfnamefont {A.~D.}\ \bibnamefont {Christianson}}, \bibinfo {author}
  {\bibfnamefont {N.}~\bibnamefont {Murai}}, \bibinfo {author} {\bibfnamefont
  {K.}~\bibnamefont {Kawashima}}, \bibinfo {author} {\bibfnamefont
  {Y.}~\bibnamefont {Yoshida}}, \bibinfo {author} {\bibfnamefont
  {H.}~\bibnamefont {Eisaki}}, \ and\ \bibinfo {author} {\bibfnamefont
  {A.}~\bibnamefont {Iyo}},\ }\href {\doibase 10.7566/JPSJ.86.093703}
  {\bibfield  {journal} {\bibinfo  {journal} {J. Phys. Soc. Jpn.}\ }\textbf
  {\bibinfo {volume} {86}},\ \bibinfo {pages} {093703} (\bibinfo {year}
  {2017})}\BibitemShut {NoStop}%
\bibitem [{\citenamefont {Biswas}\ \emph {et~al.}(2017)\citenamefont {Biswas},
  \citenamefont {Iyo}, \citenamefont {Yoshida}, \citenamefont {Eisaki},
  \citenamefont {Kawashima},\ and\ \citenamefont {Hillier}}]{Biswas2017}%
  \BibitemOpen
  \bibfield  {author} {\bibinfo {author} {\bibfnamefont {P.~K.}\ \bibnamefont
  {Biswas}}, \bibinfo {author} {\bibfnamefont {A.}~\bibnamefont {Iyo}},
  \bibinfo {author} {\bibfnamefont {Y.}~\bibnamefont {Yoshida}}, \bibinfo
  {author} {\bibfnamefont {H.}~\bibnamefont {Eisaki}}, \bibinfo {author}
  {\bibfnamefont {K.}~\bibnamefont {Kawashima}}, \ and\ \bibinfo {author}
  {\bibfnamefont {A.~D.}\ \bibnamefont {Hillier}},\ }\href {\doibase
  10.1103/PhysRevB.95.140505} {\bibfield  {journal} {\bibinfo  {journal} {Phys.
  Rev. B}\ }\textbf {\bibinfo {volume} {95}},\ \bibinfo {pages} {140505}
  (\bibinfo {year} {2017})}\BibitemShut {NoStop}%
\bibitem [{\citenamefont {Cui}\ \emph {et~al.}(2017)\citenamefont {Cui},
  \citenamefont {Ding}, \citenamefont {Meier}, \citenamefont {Bohmer},
  \citenamefont {Kong}, \citenamefont {Borisov}, \citenamefont {Lee},
  \citenamefont {Budko}, \citenamefont {Valenti}, \citenamefont {Canfield},\
  and\ \citenamefont {Furukawa}}]{Cui2017}%
  \BibitemOpen
  \bibfield  {author} {\bibinfo {author} {\bibfnamefont {J.}~\bibnamefont
  {Cui}}, \bibinfo {author} {\bibfnamefont {Q.-P.}\ \bibnamefont {Ding}},
  \bibinfo {author} {\bibfnamefont {W.~R.}\ \bibnamefont {Meier}}, \bibinfo
  {author} {\bibfnamefont {A.~E.}\ \bibnamefont {Bohmer}}, \bibinfo {author}
  {\bibfnamefont {T.}~\bibnamefont {Kong}}, \bibinfo {author} {\bibfnamefont
  {V.}~\bibnamefont {Borisov}}, \bibinfo {author} {\bibfnamefont
  {Y.}~\bibnamefont {Lee}}, \bibinfo {author} {\bibfnamefont {S.~L.}\
  \bibnamefont {Budko}}, \bibinfo {author} {\bibfnamefont {R.}~\bibnamefont
  {Valenti}}, \bibinfo {author} {\bibfnamefont {P.~C.}\ \bibnamefont
  {Canfield}}, \ and\ \bibinfo {author} {\bibfnamefont {Y.}~\bibnamefont
  {Furukawa}},\ }\href {\doibase 10.1103/PhysRevB.96.104512} {\bibfield
  {journal} {\bibinfo  {journal} {Phys. Rev. B}\ }\textbf {\bibinfo {volume}
  {96}},\ \bibinfo {pages} {104512} (\bibinfo {year} {2017})}\BibitemShut
  {NoStop}%
\bibitem [{\citenamefont {Cho}\ \emph {et~al.}(2017)\citenamefont {Cho},
  \citenamefont {Fente}, \citenamefont {Teknowijoyo}, \citenamefont {Tanatar},
  \citenamefont {Joshi}, \citenamefont {Nusran}, \citenamefont {Kong},
  \citenamefont {Meier}, \citenamefont {Kaluarachchi}, \citenamefont
  {Guillam{\'{o}}n}, \citenamefont {Suderow}, \citenamefont {Budko},
  \citenamefont {Canfield},\ and\ \citenamefont {Prozorov}}]{Cho2017}%
  \BibitemOpen
  \bibfield  {author} {\bibinfo {author} {\bibfnamefont {K.}~\bibnamefont
  {Cho}}, \bibinfo {author} {\bibfnamefont {A.}~\bibnamefont {Fente}}, \bibinfo
  {author} {\bibfnamefont {S.}~\bibnamefont {Teknowijoyo}}, \bibinfo {author}
  {\bibfnamefont {M.~A.}\ \bibnamefont {Tanatar}}, \bibinfo {author}
  {\bibfnamefont {K.~R.}\ \bibnamefont {Joshi}}, \bibinfo {author}
  {\bibfnamefont {N.~M.}\ \bibnamefont {Nusran}}, \bibinfo {author}
  {\bibfnamefont {T.}~\bibnamefont {Kong}}, \bibinfo {author} {\bibfnamefont
  {W.~R.}\ \bibnamefont {Meier}}, \bibinfo {author} {\bibfnamefont
  {U.}~\bibnamefont {Kaluarachchi}}, \bibinfo {author} {\bibfnamefont
  {I.}~\bibnamefont {Guillam{\'{o}}n}}, \bibinfo {author} {\bibfnamefont
  {H.}~\bibnamefont {Suderow}}, \bibinfo {author} {\bibfnamefont {S.~L.}\
  \bibnamefont {Budko}}, \bibinfo {author} {\bibfnamefont {P.~C.}\ \bibnamefont
  {Canfield}}, \ and\ \bibinfo {author} {\bibfnamefont {R.}~\bibnamefont
  {Prozorov}},\ }\href {\doibase 10.1103/PhysRevB.95.100502} {\bibfield
  {journal} {\bibinfo  {journal} {Phys. Rev. B}\ }\textbf {\bibinfo {volume}
  {95}},\ \bibinfo {pages} {100502} (\bibinfo {year} {2017})}\BibitemShut
  {NoStop}%
\bibitem [{\citenamefont {Suetin}\ and\ \citenamefont
  {Shein}(2017)}]{Suetin2017}%
  \BibitemOpen
  \bibfield  {author} {\bibinfo {author} {\bibfnamefont {D.~V.}\ \bibnamefont
  {Suetin}}\ and\ \bibinfo {author} {\bibfnamefont {I.~R.}\ \bibnamefont
  {Shein}},\ }\href {\doibase 10.1007/s10948-017-4404-y} {\bibfield  {journal}
  {\bibinfo  {journal} {J. Supercond. Nov. Magn.}\ }\textbf {\bibinfo {volume}
  {1}},\ \bibinfo {pages} {1} (\bibinfo {year} {2017})}\BibitemShut {NoStop}%
\bibitem [{\citenamefont {Ren}\ \emph {et~al.}(2008)\citenamefont {Ren},
  \citenamefont {Che}, \citenamefont {Dong}, \citenamefont {Yang},
  \citenamefont {Lu}, \citenamefont {Yi}, \citenamefont {Shen}, \citenamefont
  {Li}, \citenamefont {Sun}, \citenamefont {Zhou},\ and\ \citenamefont
  {Zhao}}]{Ren2008}%
  \BibitemOpen
  \bibfield  {author} {\bibinfo {author} {\bibfnamefont {Z.-A.}\ \bibnamefont
  {Ren}}, \bibinfo {author} {\bibfnamefont {G.-C.}\ \bibnamefont {Che}},
  \bibinfo {author} {\bibfnamefont {X.-L.}\ \bibnamefont {Dong}}, \bibinfo
  {author} {\bibfnamefont {J.}~\bibnamefont {Yang}}, \bibinfo {author}
  {\bibfnamefont {W.}~\bibnamefont {Lu}}, \bibinfo {author} {\bibfnamefont
  {W.}~\bibnamefont {Yi}}, \bibinfo {author} {\bibfnamefont {X.-L.}\
  \bibnamefont {Shen}}, \bibinfo {author} {\bibfnamefont {Z.-C.}\ \bibnamefont
  {Li}}, \bibinfo {author} {\bibfnamefont {L.-L.}\ \bibnamefont {Sun}},
  \bibinfo {author} {\bibfnamefont {F.}~\bibnamefont {Zhou}}, \ and\ \bibinfo
  {author} {\bibfnamefont {Z.-X.}\ \bibnamefont {Zhao}},\ }\href {\doibase
  10.1209/0295-5075/83/17002} {\bibfield  {journal} {\bibinfo  {journal} {EPL
  (Europhys. Lett.)}\ }\textbf {\bibinfo {volume} {83}},\ \bibinfo {pages}
  {17002} (\bibinfo {year} {2008})}\BibitemShut {NoStop}%
\bibitem [{\citenamefont {Wang}\ \emph
  {et~al.}(2017{\natexlab{a}})\citenamefont {Wang}, \citenamefont {Yu},
  \citenamefont {Ruan}, \citenamefont {Pan}, \citenamefont {Mu}, \citenamefont
  {Liu}, \citenamefont {Zhao}, \citenamefont {Chen},\ and\ \citenamefont
  {Ren}}]{Wang2017a}%
  \BibitemOpen
  \bibfield  {author} {\bibinfo {author} {\bibfnamefont {X.-C.}\ \bibnamefont
  {Wang}}, \bibinfo {author} {\bibfnamefont {J.}~\bibnamefont {Yu}}, \bibinfo
  {author} {\bibfnamefont {B.-B.}\ \bibnamefont {Ruan}}, \bibinfo {author}
  {\bibfnamefont {B.-J.}\ \bibnamefont {Pan}}, \bibinfo {author} {\bibfnamefont
  {Q.-G.}\ \bibnamefont {Mu}}, \bibinfo {author} {\bibfnamefont
  {T.}~\bibnamefont {Liu}}, \bibinfo {author} {\bibfnamefont {K.}~\bibnamefont
  {Zhao}}, \bibinfo {author} {\bibfnamefont {G.-F.}\ \bibnamefont {Chen}}, \
  and\ \bibinfo {author} {\bibfnamefont {Z.-A.}\ \bibnamefont {Ren}},\ }\href
  {\doibase 10.1088/0256-307X/34/7/077401} {\bibfield  {journal} {\bibinfo
  {journal} {Chin. Phys. Lett.}\ }\textbf {\bibinfo {volume} {34}},\ \bibinfo
  {pages} {077401} (\bibinfo {year} {2017}{\natexlab{a}})}\BibitemShut
  {NoStop}%
\bibitem [{\citenamefont {Bao}\ \emph {et~al.}(2015)\citenamefont {Bao},
  \citenamefont {Liu}, \citenamefont {Ma}, \citenamefont {Meng}, \citenamefont
  {Tang}, \citenamefont {Sun}, \citenamefont {Zhai}, \citenamefont {Jiang},
  \citenamefont {Bai}, \citenamefont {Feng}, \citenamefont {Xu},\ and\
  \citenamefont {Cao}}]{Bao2015}%
  \BibitemOpen
  \bibfield  {author} {\bibinfo {author} {\bibfnamefont {J.~K.}\ \bibnamefont
  {Bao}}, \bibinfo {author} {\bibfnamefont {J.~Y.}\ \bibnamefont {Liu}},
  \bibinfo {author} {\bibfnamefont {C.~W.}\ \bibnamefont {Ma}}, \bibinfo
  {author} {\bibfnamefont {Z.~H.}\ \bibnamefont {Meng}}, \bibinfo {author}
  {\bibfnamefont {Z.~T.}\ \bibnamefont {Tang}}, \bibinfo {author}
  {\bibfnamefont {Y.~L.}\ \bibnamefont {Sun}}, \bibinfo {author} {\bibfnamefont
  {H.~F.}\ \bibnamefont {Zhai}}, \bibinfo {author} {\bibfnamefont
  {H.}~\bibnamefont {Jiang}}, \bibinfo {author} {\bibfnamefont
  {H.}~\bibnamefont {Bai}}, \bibinfo {author} {\bibfnamefont {C.~M.}\
  \bibnamefont {Feng}}, \bibinfo {author} {\bibfnamefont {Z.~A.}\ \bibnamefont
  {Xu}}, \ and\ \bibinfo {author} {\bibfnamefont {G.~H.}\ \bibnamefont {Cao}},\
  }\href {\doibase 10.1103/PhysRevX.5.011013} {\bibfield  {journal} {\bibinfo
  {journal} {Phys. Rev. X}\ }\textbf {\bibinfo {volume} {5}},\ \bibinfo {pages}
  {011013} (\bibinfo {year} {2015})}\BibitemShut {NoStop}%
\bibitem [{\citenamefont {Tang}\ \emph {et~al.}(2015)\citenamefont {Tang},
  \citenamefont {Bao}, \citenamefont {Liu}, \citenamefont {Sun}, \citenamefont
  {Ablimit}, \citenamefont {Zhai}, \citenamefont {Jiang}, \citenamefont {Feng},
  \citenamefont {Xu},\ and\ \citenamefont {Cao}}]{Tang2015}%
  \BibitemOpen
  \bibfield  {author} {\bibinfo {author} {\bibfnamefont {Z.~T.}\ \bibnamefont
  {Tang}}, \bibinfo {author} {\bibfnamefont {J.~K.}\ \bibnamefont {Bao}},
  \bibinfo {author} {\bibfnamefont {Y.}~\bibnamefont {Liu}}, \bibinfo {author}
  {\bibfnamefont {Y.~L.}\ \bibnamefont {Sun}}, \bibinfo {author} {\bibfnamefont
  {A.}~\bibnamefont {Ablimit}}, \bibinfo {author} {\bibfnamefont {H.~F.}\
  \bibnamefont {Zhai}}, \bibinfo {author} {\bibfnamefont {H.}~\bibnamefont
  {Jiang}}, \bibinfo {author} {\bibfnamefont {C.~M.}\ \bibnamefont {Feng}},
  \bibinfo {author} {\bibfnamefont {Z.~A.}\ \bibnamefont {Xu}}, \ and\ \bibinfo
  {author} {\bibfnamefont {G.~H.}\ \bibnamefont {Cao}},\ }\href {\doibase
  10.1103/PhysRevB.91.020506} {\bibfield  {journal} {\bibinfo  {journal} {Phys.
  Rev. B}\ }\textbf {\bibinfo {volume} {91}},\ \bibinfo {pages} {020506}
  (\bibinfo {year} {2015})}\BibitemShut {NoStop}%
\bibitem [{\citenamefont {Smidman}\ \emph {et~al.}(2017)\citenamefont
  {Smidman}, \citenamefont {Kirschner}, \citenamefont {Adroja}, \citenamefont
  {Hillier}, \citenamefont {Lang}, \citenamefont {Wang}, \citenamefont {Cao},\
  and\ \citenamefont {Blundell}}]{Smidman2017}%
  \BibitemOpen
  \bibfield  {author} {\bibinfo {author} {\bibfnamefont {M.}~\bibnamefont
  {Smidman}}, \bibinfo {author} {\bibfnamefont {F.~K.~K.}\ \bibnamefont
  {Kirschner}}, \bibinfo {author} {\bibfnamefont {D.~T.}\ \bibnamefont
  {Adroja}}, \bibinfo {author} {\bibfnamefont {A.~D.}\ \bibnamefont {Hillier}},
  \bibinfo {author} {\bibfnamefont {F.}~\bibnamefont {Lang}}, \bibinfo {author}
  {\bibfnamefont {Z.~C.}\ \bibnamefont {Wang}}, \bibinfo {author}
  {\bibfnamefont {G.~H.}\ \bibnamefont {Cao}}, \ and\ \bibinfo {author}
  {\bibfnamefont {S.~J.}\ \bibnamefont {Blundell}},\ }\href
  {https://arxiv.org/pdf/1711.10139.pdf http://arxiv.org/abs/1711.10139} {\
  (\bibinfo {year} {2017})},\ \Eprint {http://arxiv.org/abs/1711.10139}
  {arXiv:1711.10139} \BibitemShut {NoStop}%
\bibitem [{\citenamefont {Wang}\ \emph
  {et~al.}(2016{\natexlab{a}})\citenamefont {Wang}, \citenamefont {He},
  \citenamefont {Wu}, \citenamefont {Tang}, \citenamefont {Liu}, \citenamefont
  {Ablimit}, \citenamefont {Feng},\ and\ \citenamefont {Cao}}]{Wang2016}%
  \BibitemOpen
  \bibfield  {author} {\bibinfo {author} {\bibfnamefont {Z.~C.}\ \bibnamefont
  {Wang}}, \bibinfo {author} {\bibfnamefont {C.~Y.}\ \bibnamefont {He}},
  \bibinfo {author} {\bibfnamefont {S.~Q.}\ \bibnamefont {Wu}}, \bibinfo
  {author} {\bibfnamefont {Z.~T.}\ \bibnamefont {Tang}}, \bibinfo {author}
  {\bibfnamefont {Y.}~\bibnamefont {Liu}}, \bibinfo {author} {\bibfnamefont
  {A.}~\bibnamefont {Ablimit}}, \bibinfo {author} {\bibfnamefont {C.~M.}\
  \bibnamefont {Feng}}, \ and\ \bibinfo {author} {\bibfnamefont {G.~H.}\
  \bibnamefont {Cao}},\ }\href {\doibase 10.1021/jacs.6b04538} {\bibfield
  {journal} {\bibinfo  {journal} {J. Am. Chem. Soc.}\ }\textbf {\bibinfo
  {volume} {138}},\ \bibinfo {pages} {7856} (\bibinfo {year}
  {2016}{\natexlab{a}})}\BibitemShut {NoStop}%
\bibitem [{\citenamefont {Wang}\ \emph
  {et~al.}(2017{\natexlab{b}})\citenamefont {Wang}, \citenamefont {He},
  \citenamefont {Tang}, \citenamefont {Wu},\ and\ \citenamefont
  {Cao}}]{Wang2017}%
  \BibitemOpen
  \bibfield  {author} {\bibinfo {author} {\bibfnamefont {Z.}~\bibnamefont
  {Wang}}, \bibinfo {author} {\bibfnamefont {C.}~\bibnamefont {He}}, \bibinfo
  {author} {\bibfnamefont {Z.}~\bibnamefont {Tang}}, \bibinfo {author}
  {\bibfnamefont {S.}~\bibnamefont {Wu}}, \ and\ \bibinfo {author}
  {\bibfnamefont {G.}~\bibnamefont {Cao}},\ }\href {\doibase
  10.1007/s40843-016-5150-x} {\bibfield  {journal} {\bibinfo  {journal}
  {Science China Materials}\ }\textbf {\bibinfo {volume} {60}},\ \bibinfo
  {pages} {83} (\bibinfo {year} {2017}{\natexlab{b}})}\BibitemShut {NoStop}%
\bibitem [{\citenamefont {Scalapino}(2012)}]{Scalapino2012}%
  \BibitemOpen
  \bibfield  {author} {\bibinfo {author} {\bibfnamefont {D.~J.}\ \bibnamefont
  {Scalapino}},\ }\href {\doibase 10.1103/RevModPhys.84.1383} {\bibfield
  {journal} {\bibinfo  {journal} {Rev. Mod. Phys.}\ }\textbf {\bibinfo {volume}
  {84}},\ \bibinfo {pages} {1383} (\bibinfo {year} {2012})}\BibitemShut
  {NoStop}%
\bibitem [{\citenamefont {Jiang}\ \emph {et~al.}(2013)\citenamefont {Jiang},
  \citenamefont {Sun}, \citenamefont {Xu},\ and\ \citenamefont
  {Cao}}]{Jiang2013}%
  \BibitemOpen
  \bibfield  {author} {\bibinfo {author} {\bibfnamefont {H.}~\bibnamefont
  {Jiang}}, \bibinfo {author} {\bibfnamefont {Y.-L.}\ \bibnamefont {Sun}},
  \bibinfo {author} {\bibfnamefont {Z.-A.}\ \bibnamefont {Xu}}, \ and\ \bibinfo
  {author} {\bibfnamefont {G.-H.}\ \bibnamefont {Cao}},\ }\href {\doibase
  10.1088/1674-1056/22/8/087410} {\bibfield  {journal} {\bibinfo  {journal}
  {Chin. Phys. B}\ }\textbf {\bibinfo {volume} {22}},\ \bibinfo {pages}
  {087410} (\bibinfo {year} {2013})}\BibitemShut {NoStop}%
\bibitem [{\citenamefont {Wang}\ \emph
  {et~al.}(2016{\natexlab{b}})\citenamefont {Wang}, \citenamefont {Wang},\ and\
  \citenamefont {Shi}}]{Wang2016a}%
  \BibitemOpen
  \bibfield  {author} {\bibinfo {author} {\bibfnamefont {G.}~\bibnamefont
  {Wang}}, \bibinfo {author} {\bibfnamefont {Z.}~\bibnamefont {Wang}}, \ and\
  \bibinfo {author} {\bibfnamefont {X.}~\bibnamefont {Shi}},\ }\href {\doibase
  10.1209/0295-5075/116/37003} {\bibfield  {journal} {\bibinfo  {journal} {EPL
  (Europhys. Lett.)}\ }\textbf {\bibinfo {volume} {116}},\ \bibinfo {pages}
  {37003} (\bibinfo {year} {2016}{\natexlab{b}})}\BibitemShut {NoStop}%
\bibitem [{\citenamefont {Blundell}(1999)}]{Blundell1999}%
  \BibitemOpen
  \bibfield  {author} {\bibinfo {author} {\bibfnamefont {S.~J.}\ \bibnamefont
  {Blundell}},\ }\href {\doibase 10.1080/001075199181521} {\bibfield  {journal}
  {\bibinfo  {journal} {Contemp. Phys.}\ }\textbf {\bibinfo {volume} {40}},\
  \bibinfo {pages} {175} (\bibinfo {year} {1999})}\BibitemShut {NoStop}%
\bibitem [{\citenamefont {Yaouanc}\ and\ \citenamefont {de.
  Réotier}(2011)}]{Yaouanc2011}%
  \BibitemOpen
  \bibfield  {author} {\bibinfo {author} {\bibfnamefont {A.~A.}\ \bibnamefont
  {Yaouanc}}\ and\ \bibinfo {author} {\bibfnamefont {P.~D.}\ \bibnamefont {de.
  Réotier}},\ }\href@noop {} {\emph {\bibinfo {title} {{Muon spin rotation,
  relaxation, and resonance : applications to condensed matter}}}}\ (\bibinfo
  {publisher} {Oxford University Press},\ \bibinfo {year} {2011})\ p.\ \bibinfo
  {pages} {486}\BibitemShut {NoStop}%
\bibitem [{\citenamefont {King}\ \emph {et~al.}(2013)\citenamefont {King},
  \citenamefont {de~Renzi}, \citenamefont {Cottrell}, \citenamefont {Hillier},\
  and\ \citenamefont {Cox}}]{King2013}%
  \BibitemOpen
  \bibfield  {author} {\bibinfo {author} {\bibfnamefont {P.~J.~C.}\
  \bibnamefont {King}}, \bibinfo {author} {\bibfnamefont {R.}~\bibnamefont
  {de~Renzi}}, \bibinfo {author} {\bibfnamefont {S.~P.}\ \bibnamefont
  {Cottrell}}, \bibinfo {author} {\bibfnamefont {A.~D.}\ \bibnamefont
  {Hillier}}, \ and\ \bibinfo {author} {\bibfnamefont {S.~F.~J.}\ \bibnamefont
  {Cox}},\ }\href {\doibase 10.1088/0031-8949/88/06/068502} {\bibfield
  {journal} {\bibinfo  {journal} {Phys. Scripta}\ }\textbf {\bibinfo {volume}
  {88}},\ \bibinfo {pages} {068502} (\bibinfo {year} {2013})}\BibitemShut
  {NoStop}%
\bibitem [{\citenamefont {Brewer}\ \emph {et~al.}(1986)\citenamefont {Brewer},
  \citenamefont {Kreitzman}, \citenamefont {Noakes}, \citenamefont {Ansaldo},
  \citenamefont {Harshman},\ and\ \citenamefont {Keitel}}]{Brewer1986}%
  \BibitemOpen
  \bibfield  {author} {\bibinfo {author} {\bibfnamefont {J.~H.}\ \bibnamefont
  {Brewer}}, \bibinfo {author} {\bibfnamefont {S.~R.}\ \bibnamefont
  {Kreitzman}}, \bibinfo {author} {\bibfnamefont {D.~R.}\ \bibnamefont
  {Noakes}}, \bibinfo {author} {\bibfnamefont {E.~J.}\ \bibnamefont {Ansaldo}},
  \bibinfo {author} {\bibfnamefont {D.~R.}\ \bibnamefont {Harshman}}, \ and\
  \bibinfo {author} {\bibfnamefont {R.}~\bibnamefont {Keitel}},\ }\href
  {\doibase 10.1103/PhysRevB.33.7813} {\bibfield  {journal} {\bibinfo
  {journal} {Phys. Rev. B}\ }\textbf {\bibinfo {volume} {33}},\ \bibinfo
  {pages} {7813} (\bibinfo {year} {1986})}\BibitemShut {NoStop}%
\bibitem [{\citenamefont {Harshman}\ and\ \citenamefont
  {Celio}(1986)}]{Harshman1986}%
  \BibitemOpen
  \bibfield  {author} {\bibinfo {author} {\bibfnamefont {D.~R.}\ \bibnamefont
  {Harshman}}\ and\ \bibinfo {author} {\bibfnamefont {M.}~\bibnamefont
  {Celio}},\ }\href {\doibase 10.1007/BF02394973} {\bibfield  {journal}
  {\bibinfo  {journal} {Hyperfine Interact.}\ }\textbf {\bibinfo {volume}
  {32}},\ \bibinfo {pages} {683} (\bibinfo {year} {1986})}\BibitemShut
  {NoStop}%
\bibitem [{\citenamefont {Pratt}(2000)}]{Pratt2000}%
  \BibitemOpen
  \bibfield  {author} {\bibinfo {author} {\bibfnamefont {F.~L.}\ \bibnamefont
  {Pratt}},\ }\href {\doibase 10.1016/S0921-4526(00)00328-8} {\bibfield
  {journal} {\bibinfo  {journal} {Physica B}\ }\textbf {\bibinfo {volume}
  {710}},\ \bibinfo {pages} {289} (\bibinfo {year} {2000})}\BibitemShut
  {NoStop}%
\bibitem [{\citenamefont {Brandt}(1988)}]{Brandt1988}%
  \BibitemOpen
  \bibfield  {author} {\bibinfo {author} {\bibfnamefont {E.~H.}\ \bibnamefont
  {Brandt}},\ }\href {\doibase 10.1103/PhysRevB.37.2349} {\bibfield  {journal}
  {\bibinfo  {journal} {Phys. Rev. B}\ }\textbf {\bibinfo {volume} {37}},\
  \bibinfo {pages} {2349} (\bibinfo {year} {1988})}\BibitemShut {NoStop}%
\bibitem [{\citenamefont {Brandt}(2003)}]{Brandt2003}%
  \BibitemOpen
  \bibfield  {author} {\bibinfo {author} {\bibfnamefont {E.~H.}\ \bibnamefont
  {Brandt}},\ }\href {\doibase 10.1103/PhysRevB.68.054506} {\bibfield
  {journal} {\bibinfo  {journal} {Phys. Rev. B}\ }\textbf {\bibinfo {volume}
  {68}},\ \bibinfo {pages} {054506} (\bibinfo {year} {2003})}\BibitemShut
  {NoStop}%
\bibitem [{\citenamefont {Fesenko}\ \emph {et~al.}(1991)\citenamefont
  {Fesenko}, \citenamefont {Gorbunov},\ and\ \citenamefont
  {Smilga}}]{Fesenko1991}%
  \BibitemOpen
  \bibfield  {author} {\bibinfo {author} {\bibfnamefont {V.~I.}\ \bibnamefont
  {Fesenko}}, \bibinfo {author} {\bibfnamefont {V.~N.}\ \bibnamefont
  {Gorbunov}}, \ and\ \bibinfo {author} {\bibfnamefont {V.~P.}\ \bibnamefont
  {Smilga}},\ }\href {\doibase 10.1016/0921-4534(91)90063-5} {\bibfield
  {journal} {\bibinfo  {journal} {Physica C}\ }\textbf {\bibinfo {volume}
  {176}},\ \bibinfo {pages} {551} (\bibinfo {year} {1991})}\BibitemShut
  {NoStop}%
\bibitem [{\citenamefont {Graf}\ \emph {et~al.}(1996)\citenamefont {Graf},
  \citenamefont {Yip}, \citenamefont {Sauls},\ and\ \citenamefont
  {Rainer}}]{Graf1995}%
  \BibitemOpen
  \bibfield  {author} {\bibinfo {author} {\bibfnamefont {M.~J.}\ \bibnamefont
  {Graf}}, \bibinfo {author} {\bibfnamefont {S.~K.}\ \bibnamefont {Yip}},
  \bibinfo {author} {\bibfnamefont {J.~A.}\ \bibnamefont {Sauls}}, \ and\
  \bibinfo {author} {\bibfnamefont {D.}~\bibnamefont {Rainer}},\ }\href
  {\doibase 10.1103/PhysRevB.53.15147} {\bibfield  {journal} {\bibinfo
  {journal} {Phys. Rev. B}\ }\textbf {\bibinfo {volume} {53}},\ \bibinfo
  {pages} {15147} (\bibinfo {year} {1996})}\BibitemShut {NoStop}%
\bibitem [{\citenamefont {Chandrasekhar}\ and\ \citenamefont
  {Einzel}(1993)}]{Chandrasekhar1993}%
  \BibitemOpen
  \bibfield  {author} {\bibinfo {author} {\bibfnamefont {B.~S.}\ \bibnamefont
  {Chandrasekhar}}\ and\ \bibinfo {author} {\bibfnamefont {D.}~\bibnamefont
  {Einzel}},\ }\href {\doibase 10.1002/andp.19935050604} {\bibfield  {journal}
  {\bibinfo  {journal} {Annalen der Physik}\ }\textbf {\bibinfo {volume}
  {505}},\ \bibinfo {pages} {535} (\bibinfo {year} {1993})}\BibitemShut
  {NoStop}%
\bibitem [{\citenamefont {Uemura}\ \emph {et~al.}(1991)\citenamefont {Uemura},
  \citenamefont {Le}, \citenamefont {Luke}, \citenamefont {Sternlieb},
  \citenamefont {Wu}, \citenamefont {Brewer}, \citenamefont {Riseman},
  \citenamefont {Seaman}, \citenamefont {Maple}, \citenamefont {Ishikawa},
  \citenamefont {Hinks}, \citenamefont {Jorgensen}, \citenamefont {Saito},\
  and\ \citenamefont {Yamochi}}]{Uemura1991}%
  \BibitemOpen
  \bibfield  {author} {\bibinfo {author} {\bibfnamefont {Y.~J.}\ \bibnamefont
  {Uemura}}, \bibinfo {author} {\bibfnamefont {L.~P.}\ \bibnamefont {Le}},
  \bibinfo {author} {\bibfnamefont {G.~M.}\ \bibnamefont {Luke}}, \bibinfo
  {author} {\bibfnamefont {B.~J.}\ \bibnamefont {Sternlieb}}, \bibinfo {author}
  {\bibfnamefont {W.~D.}\ \bibnamefont {Wu}}, \bibinfo {author} {\bibfnamefont
  {J.~H.}\ \bibnamefont {Brewer}}, \bibinfo {author} {\bibfnamefont {T.~M.}\
  \bibnamefont {Riseman}}, \bibinfo {author} {\bibfnamefont {C.~L.}\
  \bibnamefont {Seaman}}, \bibinfo {author} {\bibfnamefont {M.~B.}\
  \bibnamefont {Maple}}, \bibinfo {author} {\bibfnamefont {M.}~\bibnamefont
  {Ishikawa}}, \bibinfo {author} {\bibfnamefont {D.~G.}\ \bibnamefont {Hinks}},
  \bibinfo {author} {\bibfnamefont {J.~D.}\ \bibnamefont {Jorgensen}}, \bibinfo
  {author} {\bibfnamefont {G.}~\bibnamefont {Saito}}, \ and\ \bibinfo {author}
  {\bibfnamefont {H.}~\bibnamefont {Yamochi}},\ }\href {\doibase
  10.1103/PhysRevLett.66.2665} {\bibfield  {journal} {\bibinfo  {journal}
  {Phys. Rev. Lett.}\ }\textbf {\bibinfo {volume} {66}},\ \bibinfo {pages}
  {2665} (\bibinfo {year} {1991})}\BibitemShut {NoStop}%
\bibitem [{\citenamefont {Baker}\ \emph {et~al.}(2009)\citenamefont {Baker},
  \citenamefont {Lancaster}, \citenamefont {Blundell}, \citenamefont {Pratt},
  \citenamefont {Brooks},\ and\ \citenamefont {Kwon}}]{Baker2009}%
  \BibitemOpen
  \bibfield  {author} {\bibinfo {author} {\bibfnamefont {P.~J.}\ \bibnamefont
  {Baker}}, \bibinfo {author} {\bibfnamefont {T.}~\bibnamefont {Lancaster}},
  \bibinfo {author} {\bibfnamefont {S.~J.}\ \bibnamefont {Blundell}}, \bibinfo
  {author} {\bibfnamefont {F.~L.}\ \bibnamefont {Pratt}}, \bibinfo {author}
  {\bibfnamefont {M.~L.}\ \bibnamefont {Brooks}}, \ and\ \bibinfo {author}
  {\bibfnamefont {S.~J.}\ \bibnamefont {Kwon}},\ }\href {\doibase
  10.1103/PhysRevLett.102.087002} {\bibfield  {journal} {\bibinfo  {journal}
  {Phys. Rev. Lett.}\ }\textbf {\bibinfo {volume} {102}},\ \bibinfo {pages}
  {087002} (\bibinfo {year} {2009})}\BibitemShut {NoStop}%
\bibitem [{\citenamefont {Luetkens}\ \emph {et~al.}(2008)\citenamefont
  {Luetkens}, \citenamefont {Klauss}, \citenamefont {Khasanov}, \citenamefont
  {Amato}, \citenamefont {Klingeler}, \citenamefont {Hellmann}, \citenamefont
  {Leps}, \citenamefont {Kondrat}, \citenamefont {Hess}, \citenamefont
  {K{\"{o}}hler}, \citenamefont {Behr}, \citenamefont {Werner},\ and\
  \citenamefont {B{\"{u}}chner}}]{Luetkens2008}%
  \BibitemOpen
  \bibfield  {author} {\bibinfo {author} {\bibfnamefont {H.}~\bibnamefont
  {Luetkens}}, \bibinfo {author} {\bibfnamefont {H.~H.}\ \bibnamefont
  {Klauss}}, \bibinfo {author} {\bibfnamefont {R.}~\bibnamefont {Khasanov}},
  \bibinfo {author} {\bibfnamefont {A.}~\bibnamefont {Amato}}, \bibinfo
  {author} {\bibfnamefont {R.}~\bibnamefont {Klingeler}}, \bibinfo {author}
  {\bibfnamefont {I.}~\bibnamefont {Hellmann}}, \bibinfo {author}
  {\bibfnamefont {N.}~\bibnamefont {Leps}}, \bibinfo {author} {\bibfnamefont
  {A.}~\bibnamefont {Kondrat}}, \bibinfo {author} {\bibfnamefont
  {C.}~\bibnamefont {Hess}}, \bibinfo {author} {\bibfnamefont {A.}~\bibnamefont
  {K{\"{o}}hler}}, \bibinfo {author} {\bibfnamefont {G.}~\bibnamefont {Behr}},
  \bibinfo {author} {\bibfnamefont {J.}~\bibnamefont {Werner}}, \ and\ \bibinfo
  {author} {\bibfnamefont {B.}~\bibnamefont {B{\"{u}}chner}},\ }\href {\doibase
  10.1103/PhysRevLett.101.097009} {\bibfield  {journal} {\bibinfo  {journal}
  {Phys. Rev. Lett.}\ }\textbf {\bibinfo {volume} {101}},\ \bibinfo {pages}
  {097009} (\bibinfo {year} {2008})}\BibitemShut {NoStop}%
\bibitem [{\citenamefont {Takeshita}\ \emph {et~al.}(2009)\citenamefont
  {Takeshita}, \citenamefont {Kadono}, \citenamefont {Hiraishi}, \citenamefont
  {Miyazaki}, \citenamefont {Koda}, \citenamefont {Matsuishi},\ and\
  \citenamefont {Hosono}}]{Takeshita2009}%
  \BibitemOpen
  \bibfield  {author} {\bibinfo {author} {\bibfnamefont {S.}~\bibnamefont
  {Takeshita}}, \bibinfo {author} {\bibfnamefont {R.}~\bibnamefont {Kadono}},
  \bibinfo {author} {\bibfnamefont {M.}~\bibnamefont {Hiraishi}}, \bibinfo
  {author} {\bibfnamefont {M.}~\bibnamefont {Miyazaki}}, \bibinfo {author}
  {\bibfnamefont {A.}~\bibnamefont {Koda}}, \bibinfo {author} {\bibfnamefont
  {S.}~\bibnamefont {Matsuishi}}, \ and\ \bibinfo {author} {\bibfnamefont
  {H.}~\bibnamefont {Hosono}},\ }\href {\doibase
  10.1103/PhysRevLett.103.027002} {\bibfield  {journal} {\bibinfo  {journal}
  {Phys. Rev. Lett.}\ }\textbf {\bibinfo {volume} {103}},\ \bibinfo {pages}
  {027002} (\bibinfo {year} {2009})}\BibitemShut {NoStop}%
\bibitem [{\citenamefont {Lamura}\ \emph {et~al.}(2015)\citenamefont {Lamura},
  \citenamefont {Shiroka}, \citenamefont {Bonf{\`{a}}}, \citenamefont {Sanna},
  \citenamefont {{De Renzi}}, \citenamefont {Putti}, \citenamefont {Zhigadlo},
  \citenamefont {Katrych}, \citenamefont {Khasanov},\ and\ \citenamefont
  {Karpinski}}]{Lamura2015}%
  \BibitemOpen
  \bibfield  {author} {\bibinfo {author} {\bibfnamefont {G.}~\bibnamefont
  {Lamura}}, \bibinfo {author} {\bibfnamefont {T.}~\bibnamefont {Shiroka}},
  \bibinfo {author} {\bibfnamefont {P.}~\bibnamefont {Bonf{\`{a}}}}, \bibinfo
  {author} {\bibfnamefont {S.}~\bibnamefont {Sanna}}, \bibinfo {author}
  {\bibfnamefont {R.}~\bibnamefont {{De Renzi}}}, \bibinfo {author}
  {\bibfnamefont {M.}~\bibnamefont {Putti}}, \bibinfo {author} {\bibfnamefont
  {N.~D.}\ \bibnamefont {Zhigadlo}}, \bibinfo {author} {\bibfnamefont
  {S.}~\bibnamefont {Katrych}}, \bibinfo {author} {\bibfnamefont
  {R.}~\bibnamefont {Khasanov}}, \ and\ \bibinfo {author} {\bibfnamefont
  {J.}~\bibnamefont {Karpinski}},\ }\href {\doibase 10.1103/PhysRevB.91.024513}
  {\bibfield  {journal} {\bibinfo  {journal} {Phys. Rev. B}\ }\textbf {\bibinfo
  {volume} {91}},\ \bibinfo {pages} {024513} (\bibinfo {year}
  {2015})}\BibitemShut {NoStop}%
\bibitem [{\citenamefont {Khasanov}\ \emph {et~al.}(2008)\citenamefont
  {Khasanov}, \citenamefont {Luetkens}, \citenamefont {Amato}, \citenamefont
  {Klauss}, \citenamefont {Ren}, \citenamefont {Yang}, \citenamefont {Lu},\
  and\ \citenamefont {Zhao}}]{Khasanov2008}%
  \BibitemOpen
  \bibfield  {author} {\bibinfo {author} {\bibfnamefont {R.}~\bibnamefont
  {Khasanov}}, \bibinfo {author} {\bibfnamefont {H.}~\bibnamefont {Luetkens}},
  \bibinfo {author} {\bibfnamefont {A.}~\bibnamefont {Amato}}, \bibinfo
  {author} {\bibfnamefont {H.-H.}\ \bibnamefont {Klauss}}, \bibinfo {author}
  {\bibfnamefont {Z.-A.}\ \bibnamefont {Ren}}, \bibinfo {author} {\bibfnamefont
  {J.}~\bibnamefont {Yang}}, \bibinfo {author} {\bibfnamefont {W.}~\bibnamefont
  {Lu}}, \ and\ \bibinfo {author} {\bibfnamefont {Z.-X.}\ \bibnamefont
  {Zhao}},\ }\href {http://link.aps.org/doi/10.1103/PhysRevB.78.092506
  http://www.scopus.com/inward/record.url?eid=2-s2.0-53849111239{\&}partnerID=tZOtx3y1{\%}5Cnhttp://link.aps.org/doi/10.1103/PhysRevB.78.092506}
  {\bibfield  {journal} {\bibinfo  {journal} {Phys. Rev. B}\ }\textbf {\bibinfo
  {volume} {78}},\ \bibinfo {pages} {092506} (\bibinfo {year}
  {2008})}\BibitemShut {NoStop}%
\bibitem [{\citenamefont {Agterberg}\ \emph {et~al.}(2009)\citenamefont
  {Agterberg}, \citenamefont {Sigrist},\ and\ \citenamefont
  {Tsunetsugu}}]{Agterberg2009}%
  \BibitemOpen
  \bibfield  {author} {\bibinfo {author} {\bibfnamefont {D.~F.}\ \bibnamefont
  {Agterberg}}, \bibinfo {author} {\bibfnamefont {M.}~\bibnamefont {Sigrist}},
  \ and\ \bibinfo {author} {\bibfnamefont {H.}~\bibnamefont {Tsunetsugu}},\
  }\href {\doibase 10.1103/PhysRevLett.102.207004} {\bibfield  {journal}
  {\bibinfo  {journal} {Phys. Rev. Lett.}\ }\textbf {\bibinfo {volume} {102}},\
  \bibinfo {pages} {207004} (\bibinfo {year} {2009})}\BibitemShut {NoStop}%
\bibitem [{\citenamefont {Kato}\ \emph {et~al.}(2011)\citenamefont {Kato},
  \citenamefont {Batista},\ and\ \citenamefont {Vekhter}}]{Kato2011}%
  \BibitemOpen
  \bibfield  {author} {\bibinfo {author} {\bibfnamefont {Y.}~\bibnamefont
  {Kato}}, \bibinfo {author} {\bibfnamefont {C.~D.}\ \bibnamefont {Batista}}, \
  and\ \bibinfo {author} {\bibfnamefont {I.}~\bibnamefont {Vekhter}},\ }\href
  {\doibase 10.1103/PhysRevLett.107.096401} {\bibfield  {journal} {\bibinfo
  {journal} {Phys. Rev. Lett.}\ }\textbf {\bibinfo {volume} {107}},\ \bibinfo
  {pages} {096401} (\bibinfo {year} {2011})}\BibitemShut {NoStop}%
\bibitem [{\citenamefont {Berg}\ \emph {et~al.}(2009)\citenamefont {Berg},
  \citenamefont {Fradkin},\ and\ \citenamefont {Kivelson}}]{Berg2009}%
  \BibitemOpen
  \bibfield  {author} {\bibinfo {author} {\bibfnamefont {E.}~\bibnamefont
  {Berg}}, \bibinfo {author} {\bibfnamefont {E.}~\bibnamefont {Fradkin}}, \
  and\ \bibinfo {author} {\bibfnamefont {S.~A.}\ \bibnamefont {Kivelson}},\
  }\href {\doibase 10.1103/PhysRevB.79.064515} {\bibfield  {journal} {\bibinfo
  {journal} {Phys. Rev. B}\ }\textbf {\bibinfo {volume} {79}},\ \bibinfo
  {pages} {064515} (\bibinfo {year} {2009})}\BibitemShut {NoStop}%
\bibitem [{\citenamefont {Fradkin}\ \emph {et~al.}(2015)\citenamefont
  {Fradkin}, \citenamefont {Kivelson},\ and\ \citenamefont
  {Tranquada}}]{Fradkin2015}%
  \BibitemOpen
  \bibfield  {author} {\bibinfo {author} {\bibfnamefont {E.}~\bibnamefont
  {Fradkin}}, \bibinfo {author} {\bibfnamefont {S.~A.}\ \bibnamefont
  {Kivelson}}, \ and\ \bibinfo {author} {\bibfnamefont {J.~M.}\ \bibnamefont
  {Tranquada}},\ }in\ \href {\doibase 10.1103/RevModPhys.87.457} {\emph
  {\bibinfo {booktitle} {Reviews of Modern Physics}}},\ Vol.~\bibinfo {volume}
  {87}\ (\bibinfo  {publisher} {American Physical Society},\ \bibinfo {year}
  {2015})\ pp.\ \bibinfo {pages} {457--482}\BibitemShut {NoStop}%
\bibitem [{\citenamefont {Luke}\ \emph {et~al.}(1998)\citenamefont {Luke},
  \citenamefont {Fudamoto}, \citenamefont {Kojima}, \citenamefont {Larkin},
  \citenamefont {Merrin}, \citenamefont {Nachumi}, \citenamefont {Uemura},
  \citenamefont {Maeno}, \citenamefont {Mao}, \citenamefont {Mori},
  \citenamefont {Nakamura},\ and\ \citenamefont {Sigrist}}]{Luke1998}%
  \BibitemOpen
  \bibfield  {author} {\bibinfo {author} {\bibfnamefont {G.~M.}\ \bibnamefont
  {Luke}}, \bibinfo {author} {\bibfnamefont {T.}~\bibnamefont {Fudamoto}},
  \bibinfo {author} {\bibfnamefont {K.~M.}\ \bibnamefont {Kojima}}, \bibinfo
  {author} {\bibfnamefont {M.~I.}\ \bibnamefont {Larkin}}, \bibinfo {author}
  {\bibfnamefont {J.}~\bibnamefont {Merrin}}, \bibinfo {author} {\bibfnamefont
  {B.}~\bibnamefont {Nachumi}}, \bibinfo {author} {\bibfnamefont {Y.~I.}\
  \bibnamefont {Uemura}}, \bibinfo {author} {\bibfnamefont {Y.}~\bibnamefont
  {Maeno}}, \bibinfo {author} {\bibfnamefont {Z.~Q.}\ \bibnamefont {Mao}},
  \bibinfo {author} {\bibfnamefont {Y.}~\bibnamefont {Mori}}, \bibinfo {author}
  {\bibfnamefont {H.}~\bibnamefont {Nakamura}}, \ and\ \bibinfo {author}
  {\bibfnamefont {M.}~\bibnamefont {Sigrist}},\ }\href {\doibase 10.1038/29038}
  {\bibfield  {journal} {\bibinfo  {journal} {Nature (London)}\ }\textbf
  {\bibinfo {volume} {394}},\ \bibinfo {pages} {558} (\bibinfo {year}
  {1998})}\BibitemShut {NoStop}%
\bibitem [{\citenamefont {Zhang}\ \emph {et~al.}(2011)\citenamefont {Zhang},
  \citenamefont {Ye}, \citenamefont {Ge}, \citenamefont {Chen}, \citenamefont
  {Jiang}, \citenamefont {Xu}, \citenamefont {Xie},\ and\ \citenamefont
  {Feng}}]{Zhang2011}%
  \BibitemOpen
  \bibfield  {author} {\bibinfo {author} {\bibfnamefont {Y.}~\bibnamefont
  {Zhang}}, \bibinfo {author} {\bibfnamefont {Z.~R.}\ \bibnamefont {Ye}},
  \bibinfo {author} {\bibfnamefont {Q.~Q.}\ \bibnamefont {Ge}}, \bibinfo
  {author} {\bibfnamefont {F.}~\bibnamefont {Chen}}, \bibinfo {author}
  {\bibfnamefont {J.}~\bibnamefont {Jiang}}, \bibinfo {author} {\bibfnamefont
  {M.}~\bibnamefont {Xu}}, \bibinfo {author} {\bibfnamefont {B.~P.}\
  \bibnamefont {Xie}}, \ and\ \bibinfo {author} {\bibfnamefont {D.~L.}\
  \bibnamefont {Feng}},\ }\href {\doibase 10.1038/nphys2248} {\bibfield
  {journal} {\bibinfo  {journal} {Nat. Phys.}\ }\textbf {\bibinfo {volume}
  {8}},\ \bibinfo {pages} {371} (\bibinfo {year} {2011})}\BibitemShut {NoStop}%
\end{thebibliography}%

\end{document}